\documentclass[%
 aip,
rsi,%
 amsmath,amssymb,
 reprint,%
]{revtex4-1}

\usepackage{graphicx}
\usepackage{dcolumn}
\usepackage{bm}

\usepackage{listings}
\usepackage[usenames, dvipsnames]{color} 
\definecolor{mygreen}{RGB}{28,172,0} 
\definecolor{mylilas}{RGB}{170,55,241}

\begin{document}
\title{Bottom-up approach to torus bifurcation in neuron models}

\author{Huiwen Ju}
 \affiliation{Neuroscience Institute, Georgia State University
}
 
\author{Alexander B. Neiman}%
\affiliation{Department of Physics and Astronomy, Ohio University
}%

\author{Andrey L. Shilnikov}
\affiliation{Neuroscience Institute, Georgia State University
}
\affiliation{Department of Mathematics and Statistics, Georgia State University
}

\begin{abstract}
We study the quasi-periodicity phenomena occurring at the transition between tonic spiking and bursting activities in exemplary biologically plausible Hodgkin-Huxley type models of individual cells and reduced phenomenological models with slow and fast dynamics. Using the geometric slow-fast dissection and the parameter continuation approach we show that the transition is due to either the torus bifurcation or the period-doubling bifurcation of a stable periodic orbit on the 2D slow-motion manifold near a characteristic fold. We examine various torus bifurcations including stable and saddle torus-canards, resonant tori, the co-existence of nested tori and the torus breakdown leading to the onset of complex and bistable dynamics in such systems.   
\end{abstract}

\date{\today}%

\maketitle

{\bf Neurons exhibit a multiplicity of oscillatory patterns such as periodic, quasiperiodic and chaotic tonic-spiking and bursting oscillations including their mixed modes. The corresponding mathematical models fall into the class of slow-fast systems, which are characterized by the existence and interaction of diverse characteristic timescales. Study of typical transitions between these oscillatory regimes, described in terms of the bifurcation theory, is one of the trends in mathematical neuroscience. Quasiperiodic oscillations, associated with frequency locking and synchronization two close or commensurable frequencies, typically occur in coupled or periodically forced nonlinear dynamical systems. The phase space of the system with quasiperiodic oscillations contains a resonant torus, whose dimension is determined by the number of the interacting characteristic frequencies.  In slow-fast systems quasiperiodic oscillations emerge through nonlinear reciprocal interactions when the fast subsystem experiences a bottle-neck effect equating its time scale to that of the slow subsystem.  The torus breakdown, being one of the routes to onset of complex dynamics, turns out to also underlie  a typical mechanism of transition from tonic-spiking to regular or chaotic bursting in one class of slow-fast neuronal models. The study of quasiperiodicity and torus bifurcation is a challenging task that requires the aggregated use of computational approaches based on non-local bifurcation techniques. In this paper we follow the so-called bottom-up approach, starting with the highly detailed Hodgkin-Huxley type models, and ending with formal reduced models. Our goal is to demonstrate how various techniques: geometrical slow-fast dissection, parameter continuation and averaging, Poincar\'e return maps, when combined, allow us to elaborate on all fine details of the theory of torus bifurcations and breakdown in the illustrative applications. 
	
We dedicate this paper to our dear colleague and friend, Valentin Afraimovich, who passed away suddenly in February, 2018. He made many fundamental contributions to the theory of dynamical systems and bifurcations, including his co-works on quasiperodicity that set the stage for the current understanding of this phenomenon in dissipative nonlinear systems. 
}

\section{Introduction}

Deterministic modeling of oscillatory cells has been originally proposed within a framework of slow-fast dynamical systems written in a generic form:
\begin{equation}
\mathbf{x}'=F(\mathbf{x},y,\alpha), \quad  y'=\mu G(\mathbf{x},y,\alpha),
\label{abs}  
\end{equation}
where $\mathbf{x} \in R^n$, $n\ge 2$, and $y \in R^1$ represent fast and slow variables, respectively,
$|\mu| \ll 1$, and  $F,\,G$ are some smooth functions with $\alpha$ being a parameter vector. For sake of simplicity and convenience, the functions in mathematical models are often chosen so that $G(x,y)=(f(x)-y)$ to have a Jordan block in the linearization matrix, and $G$ is bi-linear in its arguments.   
 Geometric approach to study of three-dimensional (3D) models of bursting neurons was proposed and developed by J.~Rinzel \cite{Rinzel1985}. The essence for geometric understanding of neuronal dynamics is in the topology of the so-called slow-motion or critical manifolds in the phase space of the corresponding slow-fast model. The slow-fast dissection in the singular limit, $\mu=0$, freezes the slowest variable, $y$, which is then treated as a control parameter. This allows detection and parameter continuation in $y$ of the one-dimensional (1D) slow-motion manifold [given by $F=0$] of equilibria (eq), and 2D branches  of periodic orbits (PO) of the fast subsystem.

These manifolds, called quiescent M$_{\rm eq}$ and tonic-spiking M$_{\rm PO}$ in the neuroscience context, as some backbones shape the kinds of bursting activity in typical models of  individual neurons both formal and derived through the Hodgkin-Huxley formalism. Moreover, one can loosely describe the type of bursting using the bifurcations that initiate and terminate the manifolds quiescent M$_{\rm eq}$ and tonic-spiking M$_{\rm PO}$ \cite{Izhikevich2000}. For example, ``square-wave'' bursting shown in Fig.~\ref{fig:hairnemo} can be alternatively called ``fold-homoclinic'', while ``square-wave'' bursting depicted  in Fig.~\ref{fig:torusNemo} is due to the fold/saddle-node of equilibria and the fold of periodic orbits.   

The dissection method, allowing a clear geometric description and understanding the dynamics of low- and even high-dimensional models, is also helpful to predict possible transitions between activity types, e.g. tonic-spiking and bursting.  While the slow-fast dissection paradigm works well for most models with distinct timescales, it provides less or even misleading insights to understand transitions between activity patterns, which are due to reciprocal interactions of slow and fast dynamics in both subsystems even for small but finite $\mu$-values, which is typical for models of individual neurons discussed below. Such reciprocal interactions become not only non-negligible but pivotal for full understanding of both formal and biologically plausible models in question when dynamics of the fast subsystem slows down to the time scale of the slow dynamic.
%
%
\begin{figure}[t!]
\includegraphics[width=0.8\columnwidth]{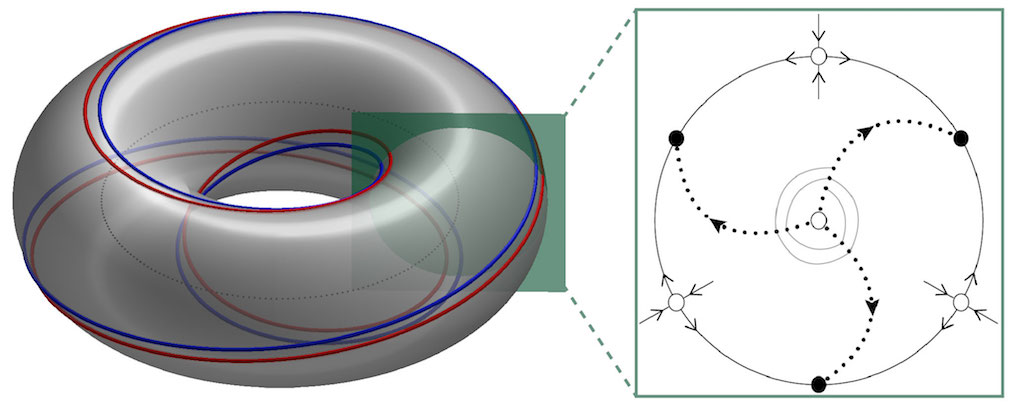}
\caption{3:1-resonant torus with a pair of periodic orbits, stable and repelling, corresponding to stable and saddle period-3 points on an invariant circle (IC) around a repelling fixed point inside, of the Poincar\'e return map on a 2D cross-section. }\label{fig:torusCartoon}
\end{figure}
This occurs universally when the fast subsystem undergoes an Andronov-Hopf (AH) bifurcation so that the rate of convergence to a critical equilibrium state is no longer exponentially fast but logarithmically slow, or when its equilibria undergo through a saddle-node/fold bifurcation with a long dwelling time due to a bottle neck effect. More complex yet very interesting cases include non-local bifurcations through which the fast subsystem possesses a stable orbit of large period that is about to become a homoclinic orbit to a saddle or to a saddle-node (SNIC), or when a pair of periodic orbits, stable and unstable, merges and vanishes in the phase space through the saddle-node  bifurcation. One of the well-known examples of such slow reciprocal interactions is related to the phenomenon of canard oscillations of small amplitudes in a relaxation oscillator model (also known as the FitzHugh-Nagumo model in computational neuroscience  \cite{izhikevich2006fitzhugh}). The canard oscillations emerge through an AH bifurcation of an equilibrium state with characteristic exponents close to zero in the norm ($\pm i \omega \sim \mu$). The shape of these small oscillations is due to the fold of the characteristic $S$-shaped branch (called the fast nullcline and given by $F=0$) of equilibria of the fast subsystem (see e.g. Fig.~\ref{fig:rinzelSN}{\bf A}-{\bf B}). The equilibrium of the full system is located at the transverse intersection of the fast nullcline with the slow ($|\mu| \ll 1$) nullcline given by $G=0$. Note that if the intersection of the slow and fast nullclines is not transverse, i.e. when $\nabla F ||  \nabla G$, then the equilibrium states in the full system bifurcate through a saddle-node. The stability of the canard orbit emerging from a preserving equilibrium state is determined solely by the fast subsystem in the singular limit, $\mu=0$, i.e., it depends on whether the AH bifurcation is sub- or super-critical. The kind of AH bifurcation is determined by the sign of the first Lyapunov coefficient $l_1$ \cite{Shilnikov2001ts}, which is 
related to the smoothness of the function $f(x)$. More specifically, given that $f'=0$ at the fold, and the constant concavity of the graph of $f(x)$ at the equilibrium state in question, then the sign of $l_1$ is related to the sign of the Taylor expansion coefficient of $f'''$ at the critical equilibrium state at the bifurcation. The choice of the function, for example $x^2$, $x^3$, $1/(x^2+1)$, etc,  whose graphs have such folds locally,  will result in either a sub- or a super-critical AH bifurcation. The same holds true when one considers the difference equations with discrete time variable, $t \in Z^+$, rather the differential one,  written in a similar form like Eqs.~(\ref{abs}) to analyze the local stability of invariant circle-canards emerging from a stable fixed point~\cite{shilnikov2004subthreshold, shilnikov2003origin}. It happens often that the sign of $l_1$ derived through the fast subsystem in the singular limit $\mu=0$ changes to the opposite in the full system even at small but finite $\mu$ values. This indicates that $l_1$ value is near zero and that the the system is close to the codimension-2 Bautin point, where $l_1=0$, in the parameter space. This indicates also that the slow-fast dissection in the singular limit is only a first order approximation for comprehensive understanding the full dynamics of a neuronal model with less desperate time scales.

The feature of the Bautin bifurcation is that its unfolding includes a saddle-node bifurcation curve corresponding to a double periodic orbit that either disappears or de-couples into stable and unstable (repelling in when $\mathbf{x} \in R^2$ and saddle in higher dimensions) orbits merge. As the result, the tonic-spiking manifold M$_{\rm PO}$ in the phase space of the system has a distinct fold (likes ones shown in Figs.~\ref{fig:hairnemo},\ref{fig:pyramid}-\ref{fig:average}, and \ref{fig:NF}-\ref{fig:nf3model} below) at a merger of stable $\mathrm{M_{PO}^S}$ and unstable, $\mathrm{M_{PO}^U}$, branches of the tonic spiking manifold, M$_{\rm PO}$. In slow-fast systems with a single slow variable, the existence of such a fold is a prerequisite, but not a guarantee, for the torus bifurcation
at the transition between tonic-spiking and bursting activity.
%

\begin{figure}[t!]
\includegraphics[width=\columnwidth]{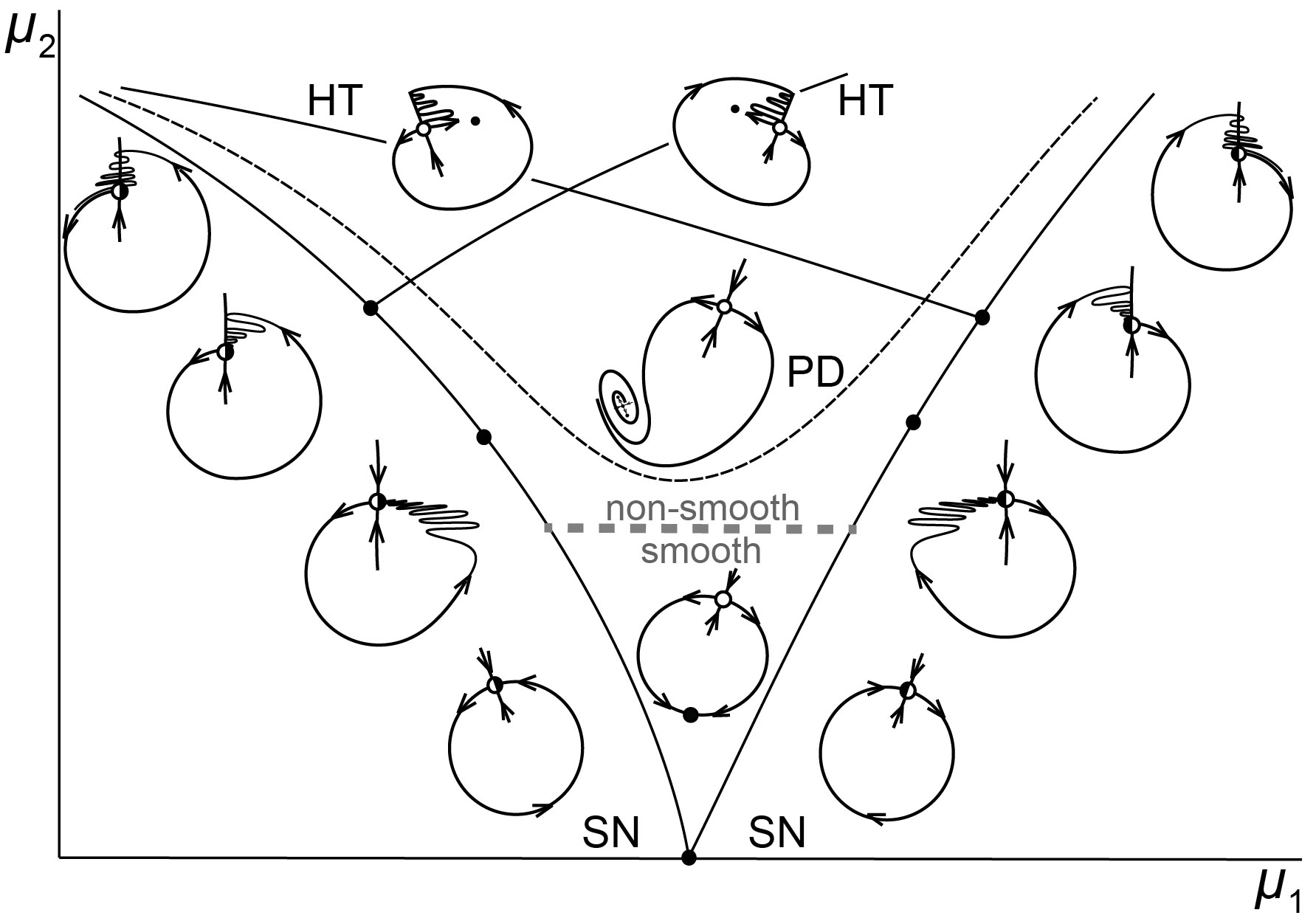}
\caption{Torus breakdown unfolding including the resonance zone that originates from the torus bifurcation curve ($\mu_2=0$) and bounded by the saddle-node (SN) bifurcation curves for fixed points (FP) on the invariant circle (IC).  At larger $\mu_2$, ICs become non-smooth first, causing the torus breakdown, followed by a period-doubling bifurcation (PD) of the stable fixed point (FP), and formations of homoclinic tangles (HT) of the saddle FP, and of the saddle-node FP on the SN-borders of the resonant zone.}
\label{fig:tongue}
\end{figure}  
A canard-torus emerges from a periodic orbit with a pair of multipliers $e^{ \pm i\phi}$ ($\phi \sim \mu$) on a unit circle.
However, the question about its stability is more challenging compared to that of a limit cycle canard emerging from an equilibrium state with a pair of complex conjugate exponents $\pm i \omega$ ($\omega \sim \mu$), because the torus stability cannot be determined {\em a priori} by a local stability analysis. En route to bursting the stable periodic orbit, associated with tonic spiking activity $\mathrm{M_{PO}^S}$, may loose its stability via a period doubling bifurcation (single multiplier $e^{ i\pi}=-1$), rather than though a torus bifurcation whether it is super- or subcritical, giving rise a cascade of period doubling  \cite{Cymbalyuk2005, Shilnikov2012}. 

In the case of the torus bifurcation, the emerging stable or unstable (repelling in $R^3$ or saddle in $R^{4+}$) torus can be resonant or ergodic, depending on the angle $\phi$ of its multipliers on an unit circle.
For example, strong resonances such as 1:4 and 1:3 occur when $\phi=\pi/2$ and $\phi=2\pi/3$, respectively. The torus is born ergodic when $\phi$ is not commensurable with $\pi$.
The example of a resonant~3:1 torus with a rational Poincar\'e winding number 3/1 is shown in Fig.~\ref{fig:torusCartoon}. In simple terms this means that there is a pair of periodic orbits, stable and unstable, on the 2D torus that correspond to a pair of periodic orbits of period-3 on the invariant circle in some local cross-section transverse to the torus. The winding number remains constant within a synchronization zone (see Fig.~\ref{fig:torusCartoon}), also known as an Arnold tongue, and takes varying irrational values outside of the synchronization zone. In this example of the resonant torus, the frequency locking ratio 3/1 means that the ``horizontal''  oscillations are three times faster than the ``vertical'' slow-components due to the timescales of $\mathbf{x}$ and $y$-variables of Eqs.~(\ref{abs}).

     Andronov and Vitt \cite{Zur1930} were first to study synchronization or phase-locking in the periodically forced van der Pol equation \cite{shilnikov2004some}.      
  They showed that the wedge-like shape of each resonant zone is bounded by the saddle-node bifurcation (SNIC) curves in the parameter plane so that having crossed one, the resonant torus becomes ergodic with everywhere dense covering (trajectory) on it. Later, Afraimovich and Shilnikov \cite{afraimovich1974small, afraimovich1991invariant, Annulus1977, some1974} proposed a phenomenological scenario of torus breakdown. It is illustrated in Fig.~\ref{fig:tongue},  which  depicts pivotal evolution stages of the invariant circle of a 2D Poincar\'e return map and fixed points on it inside and outside of the resonant zone in the parameter plane. Here $\mu_2=0$ corresponds to the torus bifurcation and $\mu_1$ controls the angle of the multipliers of the bifurcating periodic orbit within the range from zero through $\pi$ ($e^{ \pm i\phi}$). At small $\mu_2$ values, outside the resonant zone, the invariant circle (IC) (read the 2D torus) is initially smooth and rounded. It possesses a saddle-node fixed point (FP) on SN-curves so that the closure of its  1D unstable set constitutes the IC. Inside the resonant zone, there are two FPs, stable and saddle, on the IC. As $\mu_2$ is increased the IC becomes ``non-smooth'' so that the stable FP later undergoes a period-doubling bifurcation (PD curves) that breaks the torus down. Next, the unstable sets of the saddle-node or saddle FP touches (HT-curves) and crosses its stable sets and creates homoclinic tangles giving rise to complex chaotic dynamics in the system. The Reader is welcome to consult with the review~\cite{shilnikov2004some} and the reference texts~\cite{Shilnikov2001ts} for more details about the torus breakdown.     

\begin{figure*}[ht!]
	\includegraphics[width=.99\textwidth]{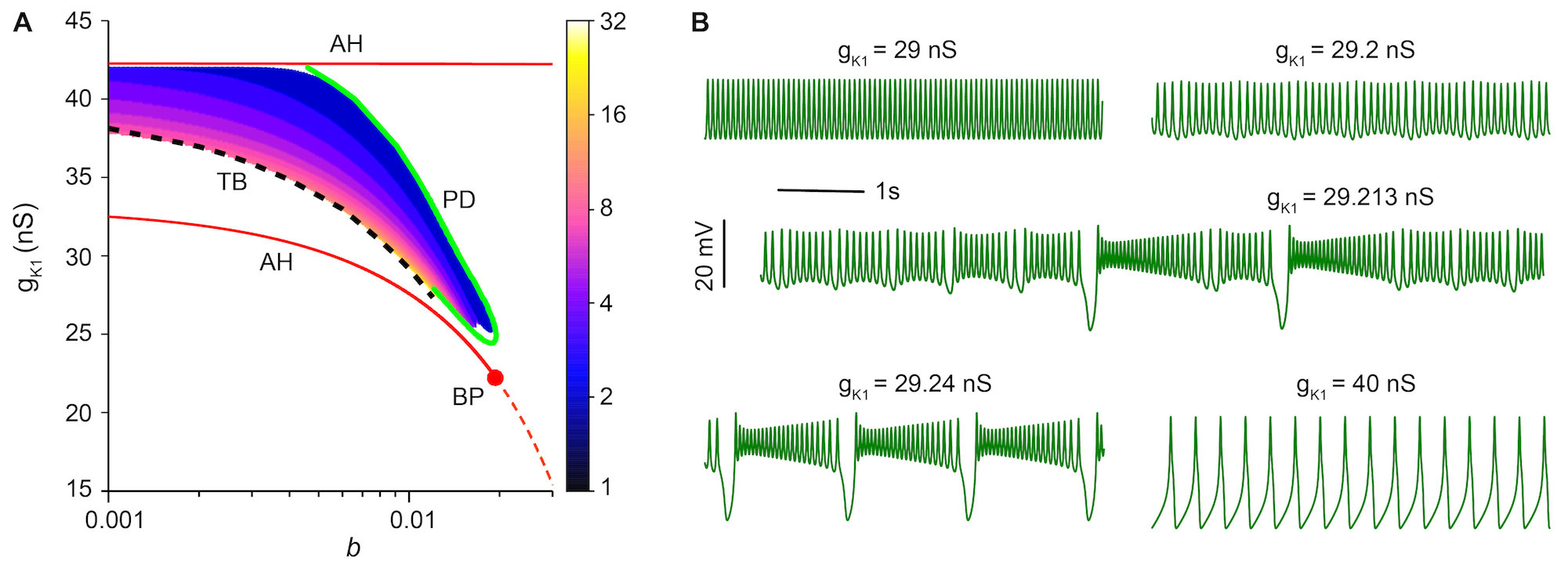}
	\caption{Dynamical regimes of the hair cell model. 
		$\mathbf{A}$ $(b,\, g_{\rm K1})$ bifurcation diagram of the hair cell model is superimposed with a color-coded map indicating the spike number per burst. It includes Andronov-Hopf (red) curves of supercritical (solid line) and subcritical (dashed line) bifurcations including a codimension-2 Bautin point (filled circle, BP) that bounds the region of oscillatory activity of the hair cell. Black dashed line corresponds to the torus bifurcation (TB), and solid green line corresponds to a period-doubling (PD) bifurcation of the periodic orbit (PO).  
		$\mathbf{B}$ Gallery of voltage traces for the indicated increasing $g_\mathrm{K1}$-values  at fixed  $b=0.01$.
	}
	\label{fig:hair1}
\end{figure*}

In-depth understanding of the universal  mechanisms of transitions between oscillatory activity patterns in single neuron models is a challenge for the theory of bifurcations. It requires the use of wide range of advanced apparatus of the bifurcation theory, ranging from the blue sky catastrophe to various homoclinic bifurcations of saddle orbits and equlibria  \cite{Gaspard1987,Gaspard1992a,Terman1992,Feudel2000,Shilnikov2001ts,Belykh2000,Shilnikov2005,Shilnikov2005b,Cymbalyuk2005,Mmo2005,Channell2007,Channell2007a,Shilnikov2008a,Shilnikov2012}.

In what follows we will present several examples of biologically plausible and formal mathematical slow-fast models to demonstrate the universality and complexity of the the torus bifurcations near the distinct fold on the tonic-spiking manifold. We use the bottom-up approach, starting with highly detailed models and ending with simple toy models, all featuring same generic properties.  These bifurcations can result in the onset of stable, repelling and saddle tori as well as their coexistence in the phase space leading to bi-stability of tonic spiking and bursting activities. We  discuss the mechanisms of torus breakdowns through various resonances, and conditions for period-doubling bifurcations alternatively occurring on the same fold. We note the torus-canards have become a new trend in mathematical neuroscience with a focus on transition between tonic-spiking and bursting in various slow-fast models of individual neurons see and references therein \cite{2012showcase, benes2011elementary, Wojcik2011, vo2017generic, Kramer2008a, alexander2011spontaneous, Shilnikov2005b, shilnikov2012complete}.

\section{A model for spontaneous voltage oscillations in hair cell}
Our first example is a  detailed Hodgkin-Huxley type model, whose 12 ODEs describe spontaneous dynamics of membrane potential of a hair cell. 
Hair cells are peripheral receptors which transform mechanical stimuli into electrical signals in the senses of hearing and balance of vertebrates.
These sensors rely on nonlinear active processes to achieve astounding sensitivity, selectivity and dynamical range  \cite{hudspeth2014integrating}.   
In particular, it was proposed that ciliary bundles of hair cells may self-tune to operate on the verge of (degenerate) AH bifurcations causing a sharpened selectivity to weak periodic mechanical forcing \cite{camalet2000auditory,equiluz2000}.

In amphibians  some hair cells demonstrate various spontaneous oscillatory activities. In particular, several experimental studies have documented various oscillation regimes of membrane potential in the frog sacculus \cite{Catacuzzeno2003,Rutherford2009}.
Although physiological implications of the membrane potential oscillations on sensory function of hair cells are not clear, experimental study \cite{Rutherford2009} showed that spontaneous voltage oscillations drive sensory neurons resulting in periodic firing. 
Furthemore, the membrane potential of the hair cell affect the dynamics of its apical hair bundle compartment, as documented experimentally 
\cite{ meenderink2015voltage} and in modeling work \cite{han2010spontaneous,khamesian2017effect}.
Modeling work \cite{montgomery2007amplification,han2010spontaneous,amro2014effect} predicted that oscillatory dynamics of the membrane potential  may lead to improved sensory performance of the hair cell.

\subsection{Hair cell model}
The model is based on several experimental studies of basolateral ionic currents in saccular hair cells in bullfrog \cite{Hudspeth1988a,Catacuzzeno2003,Catacuzzeno2004,Rutherford2009} and is a modification of a model developed in \cite{Catacuzzeno2004}. It includes 6 ion currents plus a leak current with corresponding equations for kinetics of ion channels and for dynamics of Ca$^{2+}$ concentration, resulting in a Hodgkin-Huxley type system of 12 coupled nonlinear ordinary differential equations.
A detailed description of the model is available in an open access publication \cite{alexander2011spontaneous} and a code for the right hand sides of corresponding ODEs is provided in the Appendix .
\begin{figure*}[t!]
	\includegraphics[width=.8\textwidth]{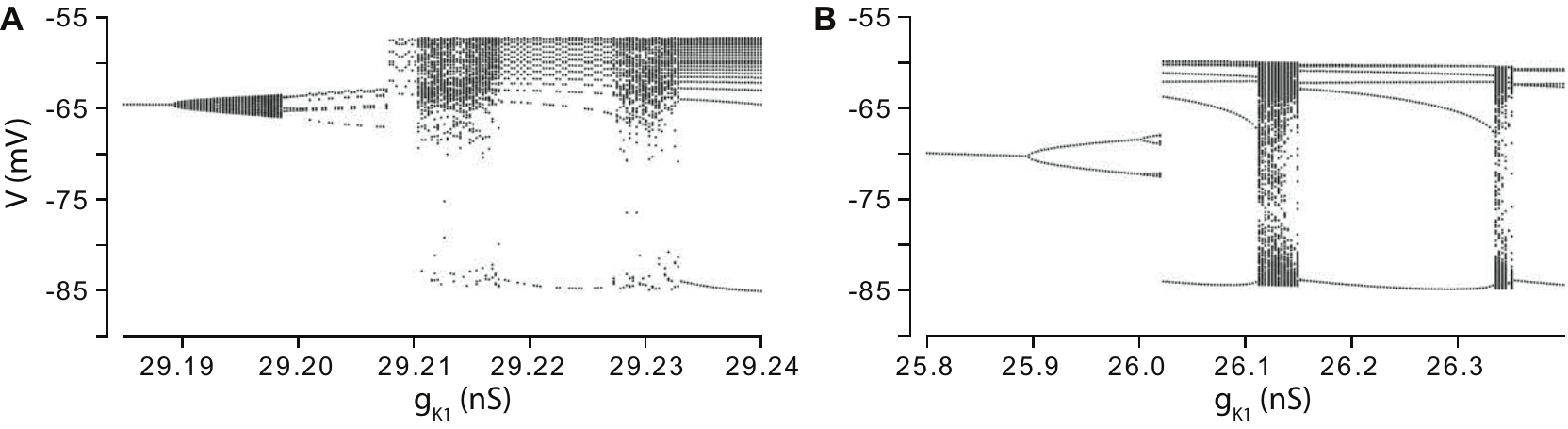}
	\caption{$\mathbf{A}$ Bifurcation diagram representing the $g_{\rm K1}$-parameter sweep of $V_{\rm min}$-values reveals the stability loss of the tonic-spiking periodic orbit through a torus bifurcation en route to bursting at level $b=0.01$. Note that the ergodic torus becomes resonant before its breakdown into large-amplitude bursts. $\mathbf{B}$ Diagram showing the cascade of period-doubling bifurcations of tonic-spiking orbits transitioning  to bursting at level $b=0.015$.}
	\label{fig:hair2}
\end{figure*}

The membrane potential dynamics in the hair cell model is described by this equation:
\begin{equation}
C_\mathrm{m}V' = -I_\mathrm{K1}-I_\mathrm{h}-I_\mathrm{DRK}-I_\mathrm{Ca}-I_\mathrm{BKS}-I_\mathrm{BKT}-I_\mathrm{L}.
\label{eq:hair} 
\end{equation}
These six ionic currents can be grouped according to their activation patterns.
The K$^{+}$ inward-rectifier ($I_\mathrm{K1}$) and the cation h-type 
($I_\mathrm{h}$) currents are hyperpolarization activated. The rest four currents are depolarization activated: the voltage activated K$^{+}$ direct-rectifier ($I_\mathrm{DRK}$)
and Ca$^{2+}$ ($I_\mathrm{Ca}$) currents; the calcium-activated K$^{+}$ steady 
($I_\mathrm{BKS}$) and transient ($I_\mathrm{BKT}$) currents. Finally, the leak current 
($I_\mathrm{L}$) stands for the mechano-electrical transduction from the hair bundle compartment of the cell.
Previous computational study \cite{alexander2011spontaneous} showed that a convenient choice of control parameters is the maximal conductance
of K1-current, $g_{\rm K1}$, and the strength of the Ca$^{2+}$-activated potassium currents, $b$.

Experimental work showed that depolarization activated currents are responsible for the so-called phenomenon of electrical resonance, whereby the hair cell shows fast oscillatory responses when knocked by an external current pulse  \cite{Crawford1981c,Art1986,Lewis1988,Hudspeth1988,Hudspeth1988a}. Furthermore,
a self-sustained voltage oscillations were observed experimentally and in a model which contained the voltage-activated Ca$^{2+}$ and the Ca$^{2+}$-activated potassium currents \cite{Ospeck2001}. On the other hand, hyperpolarization activated currents give rise to a slow (3--4~Hz) large-amplitude oscillations of the membrane potential, owing to slow kinetics of $I_\mathrm{h}$ current \cite{Catacuzzeno2004}. Taken together, the membrane potential of the frog  hair cells show diverse patterns of oscillatory activity documented experimentally \cite{Rutherford2009}, and well captured by the model \cite{alexander2011spontaneous}. In particular, the two distinct oscillatory mechanisms mentioned above may lead to quasi-periodic oscillations, with a torus bifurcation and torus breakdown \cite{alexander2011spontaneous}, which will be studied here in details. 

\subsection{Quasi-periodicity \& period-doubling pathways}
To unravel the dynamical mechanisms behind transition to quasiperiodic oscillations
we built a bifurcation diagram of the model for two control parameters,  $g_{\rm K1}$ and $b$. The diagram shown in Fig.~\ref{fig:hair1} was constructed using the combined techniques of parameter sweeping and continuation.

In  Fig.~\ref{fig:hair1}$\mathbf{A}$, the AH bifurcation (red curve)  demarcates the region of the quiescence state from the oscillatory dynamics. The solid and dashed segments of the AH line correspond to super- and subcritical  AH  bifurcation. The criticality of the bifurcation means that either a stable or a saddle periodic orbit emerges and collapses into the stable equilibrium state, after the corresponding segment of the AH-curve is crossed. The point, labeled BP, is where the criticality of the AH-bifurcation changes. This bifurcation was first studied by N.~Bautin \cite{Shilnikov2001ts} who found that its unfolding includes another curve corresponding to a double, saddle-node periodic orbit around the equilibrium state. The orbit results from a coalescence of the stable and saddle ones, that have emerged from a stable focus through  super- and sub-critical AH bifurcations, respectively, upon crossing the solid and dashed branches of the (red) curve, AH, in the diagram in Fig.~\ref{fig:hair1}{\bf A}. 

\begin{figure}[b!] 
\includegraphics[width=\columnwidth]{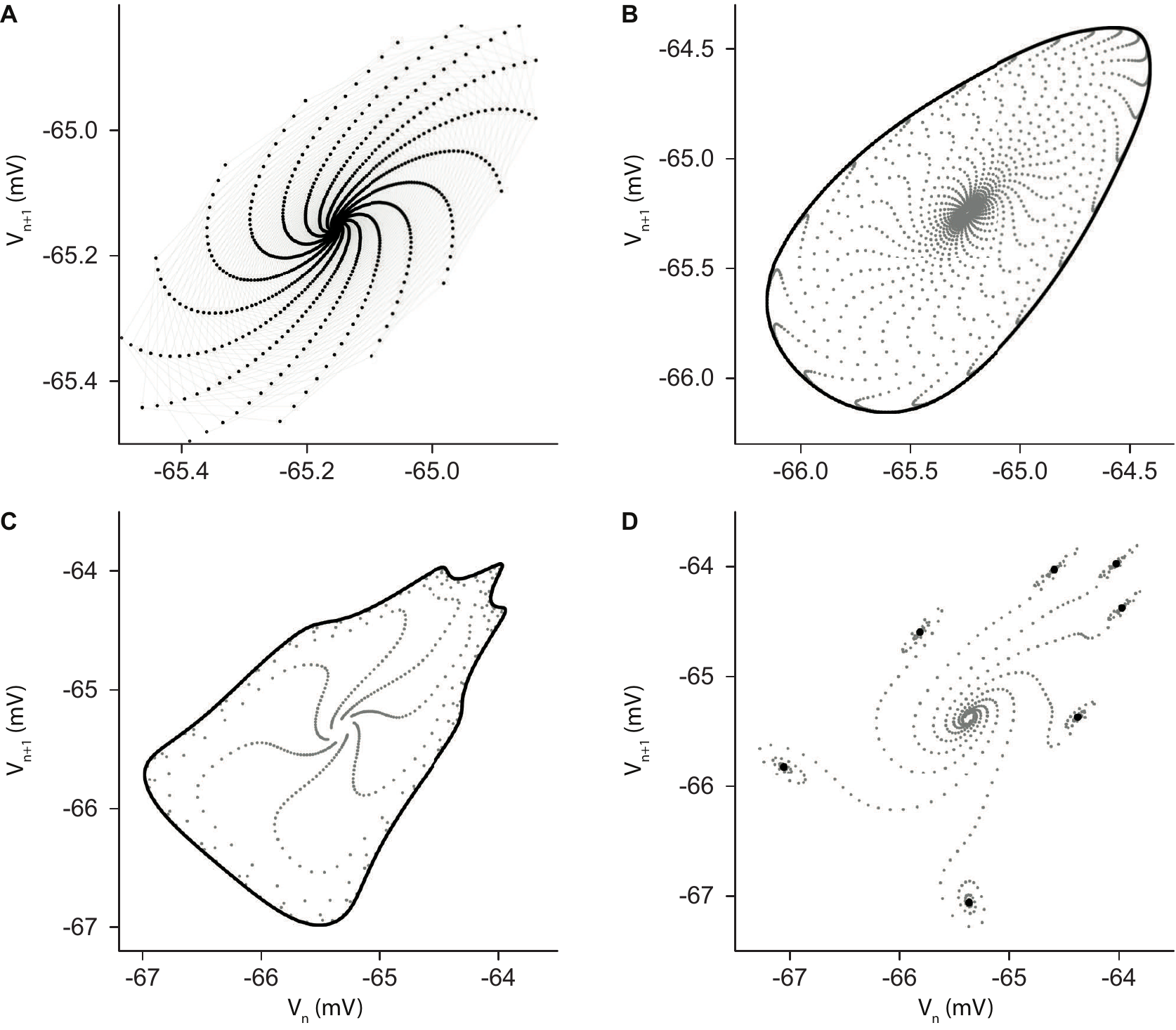}
\caption{Poincar\'e return maps, $V_{\rm min}^{(n)} \to V_{\rm min}^{(n+1)}$, depicting four pivotal stages of the torus onset and breakdown on the pathway  $b=0.011$. 
$\mathbf{A}$ Convergence to a spiraling (resonant) fixed point (FP) corresponding to a tonic-spiking orbit at $g_\mathrm{K1}=28.335$~nS.
$\mathbf{B}$ Stable smooth IC corresponding to a stable ergodic torus at  $g_\mathrm{K1}=28.345$~nS  with a repelling FP inside.
$\mathbf{C}$ A non-smooth (distorted) IC for the yet ergodic torus at $g_\mathrm{K1}=28.355$~nS.
$\mathbf{D}$ A stable 7-period orbit at  $g_\mathrm{K1}=28.3605$~nS.}
\label{fig:hairtsprocess}
\end{figure}
A synonym of the saddle-node bifurcation of POs is a fold where two branches, foliated by stable and saddle periodic orbits, merge. Such a fold occurs on a 2D tonic-spiking manifold $\mathrm{M_{PO}}$ that is traced down by the periodic orbits of the full system  using the parameter continuation (more about this method in \cite{Shilnikov2012}). The notion of the fold is imperative here, as it is where a round tonic-spiking periodic orbit loses stability and tonic spiking transforms into bursting activity though either a series of period doubling bifurcations or through a torus bifurcation. 

There is a novel dynamic feature that makes the hair cell model stand out in the list of conventional models of bursters. Namely,  it is due to the fold on the tonic spiking manifold, $\mathrm{M_{PO}}$, that results from the saddle-node bifurcation of periodic orbits, originating in turn from the codimension-2 Bautin bifurcation on the AH curve in the diagram in Fig.~\ref{fig:hair1}. The fold {\em per se} is no dynamic feature in slow-fast systems, unless the stable periodic orbit for tonic oscillations slides down closer to the fold to lose the stability to the 2D torus \cite{Wojcik2011}. This mechanism, first described in \cite{Cymbalyuk2005}, happens to occur typically in slow-fast systems with such folds \cite{kuzn}, particularly in neuronal models \cite{Wojcik2011, 2012showcase} including the reduced model of Purkinje cells \cite{Kramer2008a} below.   

A limit cycle born through the AH bifurcation loses its stability either through a torus bifurcation (TB, dashed black line in Fig.~\ref{fig:hair1}$\mathbf{A}$) or through a period-doubling bifurcation (PD, solid green line in Fig.~\ref{fig:hair1}$\mathbf{A}$). A region bounded by these lines is characterized by various bursting patterns and by period-adding bifurcations. The colormap in Fig.~\ref{fig:hair1}$\mathbf{A}$, showing the number of spikes per burst, indicates that the number of spikes increases towards the torus bifurcation. To the right of the PD-curve, 
i.e. for $0.02< b < 1$, the model produces robustly tonic oscillations \cite{alexander2011spontaneous}, which underlines stabilizing physiological role of the Ca$^{2+}$-activated potassium currents ($I_\mathrm{BKS,BKT}$) \cite{Rutherford2009}. 
\begin{figure}[t!]
	\includegraphics[width=\columnwidth]{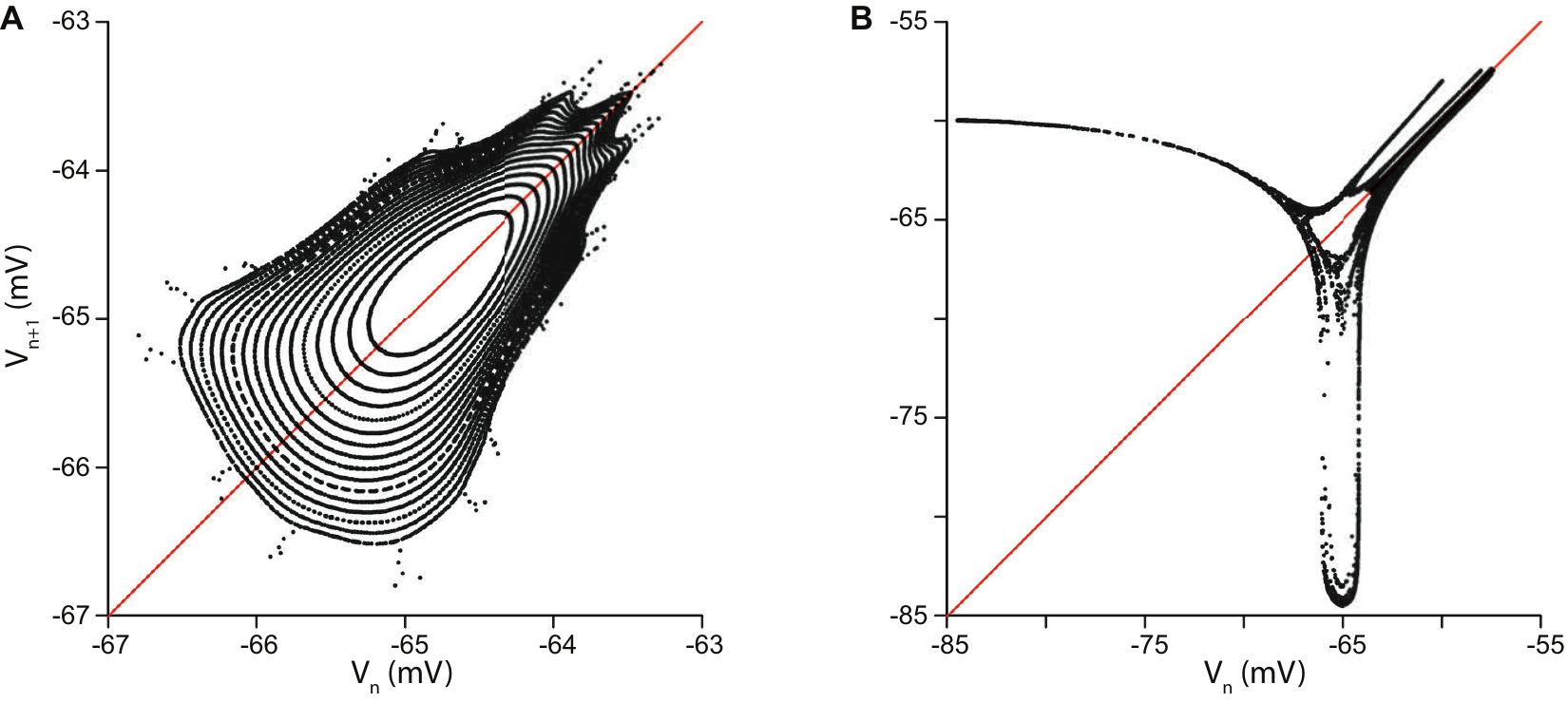}
	\caption{Poincar\'e return maps, $V_{\rm min}^{(n)} \to V_{\rm min}^{(n+1)}$, for the consecutive $V_{\rm min}$-values in voltage traces generated by the hair cell model. $\mathbf{A}$ Evolution of stable invariant circles (IC) from ergodic to resonant with further non-smooth torus breakdown as the $g_\mathrm{K1}$  parameter is increased from 29.185 through 29.2073~nS. $\mathbf{B}$ Chaotic bursting after the torus breakdown at $g_\mathrm{K1}=29.213$~nS. 
		The flat, stabilizing section of the map corresponds to the hyperpolarized quiescence, while multiple sharp folds reveals a ghost of the non-smooth IC in the depolarized range.}
	\label{fig:hairpoinc}
\end{figure}

Next we investigate the torus bifurcation and the torus breakdown using the slow-fast manifolds and Poincar\'e return maps on a route from tonic spiking to bursting. We also examine the transition to bursting on another route featuring the period-doubling bifurcation, and contrast the corresponding return maps in both cases. The Poincar\'e return map is well suited for exploration of complex oscillations,  revealing  the mechanisms of underlying bifurcations and their precursors.
Since the voltage is the only observable variable in many experimental setups, we will employ the Poincar\'e return map $T$ defined on consecutive minima of the membrane voltage trace as follows: $T: V_{\rm min}^{(n)} \to V_{\rm min}^{(n+1)}$. In the case of a sparse map (insufficiency of various points $V_{\rm min}$), e.g.,  for periodic or weak chaotic bursting, a small random perturbation can be added to the model, resulting in  variability sufficient to obtain densely populated iterates, revealing some key dynamical features. We point out that sensitivity of solutions to external perturbations is a common feature of  slow-fast systems like neuronal models \cite{Channell2009}.  

\begin{figure}[t!]
\includegraphics[width=\columnwidth]{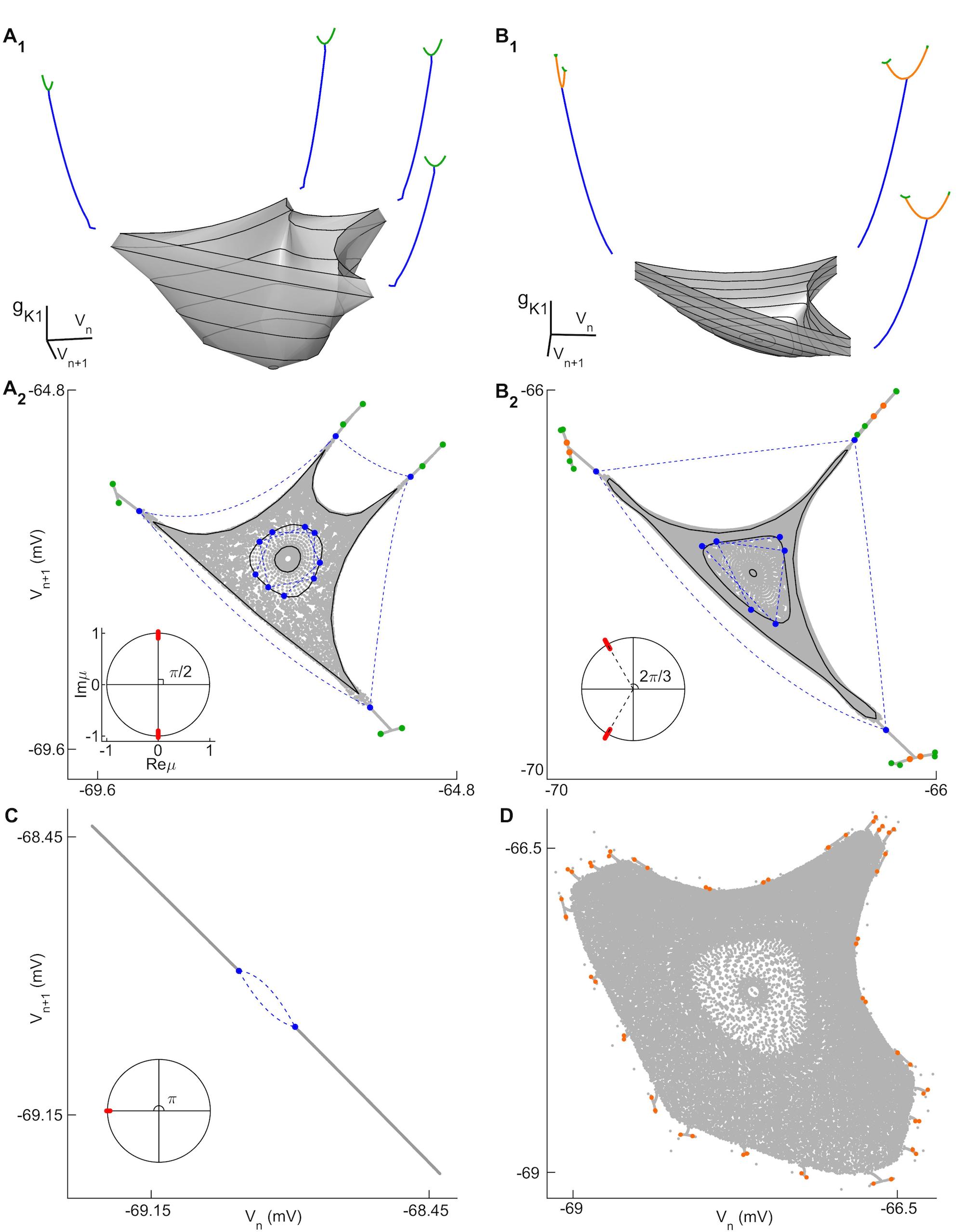}
\caption{Poincar\'e return maps, $T: V_{\rm min}^{(n)} \to V_{\rm min}^{(n+1)}$, depicting strong-resonant ICs (tori) as $g_\mathrm{K1}$-parameter increases at various fixed $b$-values. Insets $\mathbf{A}$-$\mathbf{C}$ illustrate the evolution of initially ergodic IC as it becomes strongly resonant, 1:4, 1:3 and 1:2 at $b = 0.0125, 0.0128$ and $0.013$, resp. $\mathbf{D}$ Weak resonant case at $b = 0.0127$. Positions of the Floquet multipliers [of the central FP] on a unit circle determine the resonance number. $\mathbf{A_1}$--$\mathbf{B_1}$ 3D extended evolution of the ICs with increasing  $g_\mathrm{K1}$; grey dots represent $V_{\rm min}$-values extracted from voltage traces; black contour lines represent the ICs of the return maps at specific $g_\mathrm{K1}$-values; blue dots are connected to show the iteration order; green and orange branches reveal the PD bifurcations of resonant 3- and 4-period orbits.}
\label{fig:alien}
\end{figure}

The described two scenarios of the transition from tonic-spiking to bursting (through the torus or period-doubling bifurcations) can be differentiated with the aid of  the Poincar\'e maps for consecutive minimal values of the membrane voltage traces shown in Fig.~\ref{fig:hair2}.

\subsection{From tonic spiking to bursting through quasi-periodicity} 
Let us first examine the torus bifurcation scenario, as $g_{\rm K1}$ increases with a fixed value of $b=0.01$. Typical evolution of the voltage traces is shown in Fig.~\ref{fig:hair1}$\mathbf{B}$. As the parameter $g_\mathrm{K1}$ increases a small-amplitude periodic oscillations become quasi-periodic (slow amplitude modulation) through the supercritical (stable) torus bifurcation. When the torus brakes down, it gives rise to large-amplitude bursting oscillations which may be chaotic or regular in alternation. This is illustrated in the bifurcation diagram of Fig.~\ref{fig:hair2}$\mathbf{A}$, which was built 
using the Poincar\'e return map of consecutive voltage minima. In this diagram, a single point for a given parameter value corresponds to a stable periodic orbit. Several points per a single parameter value may correspond to different regimes. Dense points above $-70$~mV correspond to ergodic tori, whereas multiple extended branches correspond to stable periodic orbit(s) on the resonant torus exiting within its resonant zone. Points below $-70$~mV, refer to bursting orbits whereby the cell becomes strongly hyperpolarized. Bursting orbits are stable when their several branches in the bifurcation diagram are continued without interruption within a stability window, followed by regions of chaotic bursting.

One can see from the bifurcation diagram in Fig.~\ref{fig:hair2}$\mathbf{A}$ that the stable periodic orbit, earlier emerging through the supercritical AH bifurcation, loses its stability when $g_{\rm K1}$ is increased through $29.19$~nS.  This stability loss gives rise to the emergence of a 2D stable torus, first ergodic, as a widening parabola-shaped solid region in the diagram suggests. In what follows, the breakdown of the torus comes along with the generic scenario lines proposed by Afraimovich and Shilnikov \cite{AS1983}. The torus becomes resonant as $g_{\rm K1}$ increases through a synchronization boundary (of an Arnold tongue) \cite{Shilnikov2001arnold, shilnikov2004some}. This assertion is supported by the occurrence of several branches corresponding to the local minima of a stable periodic orbit coexisting with a saddle orbit on the torus after a saddle-node bifurcation at $g_{\rm K1} \simeq 29.199$~nS. Further increase  of the control parameter causes the torus breakdown, which leads to bursting oscillations with modulatory (quasi-periodic) tonic spiking episodes interrupted by slow hyperpolarized quiescent periods (Fig.~\ref{fig:hair1}$\mathbf{B}$ at $g_{\rm K1}=29.213$~nS). Fig.~\ref{fig:hair2}$\mathbf{A}$ also reveals a number of spike adding bifurcation of bursting. The cascade of spike adding ends on the upper branch of the green bifurcation curve, that corresponds to a reverse period-doubling bifurcation (Fig.~\ref{fig:hair1}$\mathbf{A}$). To the right of the curve, the hair cell model produces stably a large-amplitude voltage oscillations (Fig.~\ref{fig:hair1}$\mathbf{B}$ at $g_{\rm K1}=40$~nS). The black dashed curve in Fig.~\ref{fig:hair1}$\mathbf{A}$ corresponds to the torus formation and therefore quasi-periodic tonic spiking oscillations in the model.  
%

Figures~\ref{fig:hairtsprocess} and \ref{fig:hairpoinc} represent the Poincar\'e return maps depicting the characteristic stages of the torus formation and breakdown. In Fig.~\ref{fig:hairpoinc}$\mathbf{A}$, a family of invariant circles (ICs) corresponds to the growing of the first smooth and later non-smooth ergodic stable IC emerging from a tonic-spiking periodic orbit, as the latter has lost ``its skin'' to the 2D torus (Fig.~\ref{fig:hairtsprocess}$\mathbf{A}$ and $\mathbf{B}$). As $g_{\rm K1}$ increases, the torus becomes resonant with a stable periodic orbit on it. The degree of the resonance can be evaluated by the number (here 7) of periodic points of the stable orbit, which equals the number of the  "sleeves" originating from an unstable FP towards the non-smooth invariant curve. This agrees well with the theory of torus breakdown that causes the onset of complex chaotic dynamics. In essence, the torus breakdown begins with the emergence of two periodic orbits, stable and saddle, on the torus when it becomes resonant through a saddle-node bifurcation. Later the torus becomes non-smooth after the stable and unstable manifolds (sets) of the saddle orbit start making wiggles transitioning to homoclinic tangles with the stability loss of the counterpart periodic orbit through a period-doubling bifurcation. This bifurcation can be identified through end-branching on the resonant traces left by the periodic-point as the control parameter $g_{\rm K1}$ is increased. One can observe from Fig.~\ref{fig:hairtsprocess}$\mathbf{D}$ that the stable 7-period orbit has a pair of leading complex multipliers, which indicates the proximity of period-doubling bifurcations (green branches indicated in  Fig.~\ref{fig:alien} below).

We note that bursting can already co-exist with the stable torus in the hair cell model, so the torus breakdown gives rise to bursting, first regular and then chaotic at $g_{\rm K1}=29.213$~nS. 
The corresponding return map is shown in Fig.~\ref{fig:hairpoinc}$\mathbf{B}$. It has a distinct shape, so characteristic for the square wave bursters \cite{Channell2007a,Channell2009}. The map has three components:  the transient ``toroidal'' section rivaling a ghost of the disappeared non-smooth IC around $\mathrm{V_{min}}=-65$mV, a  dropping-down middle section and a flat (contracting) section corresponding to the slow hyperpolarized quiescence phase of bursting reaching $\mathrm{V_{min}}=-85$mV. The voltage trace generating this map at $g_{\rm K1}$ is depicted in Fig.~\ref{fig:hair1}$\mathbf{B}$ at $g_{\rm K1}=29.213$~nS; it demonstrates slowly modulated chaotic tonic-spiking phases alternating  with hyperpolarized voltage sags.   

\begin{figure}[t!]
	\includegraphics[width=\columnwidth]{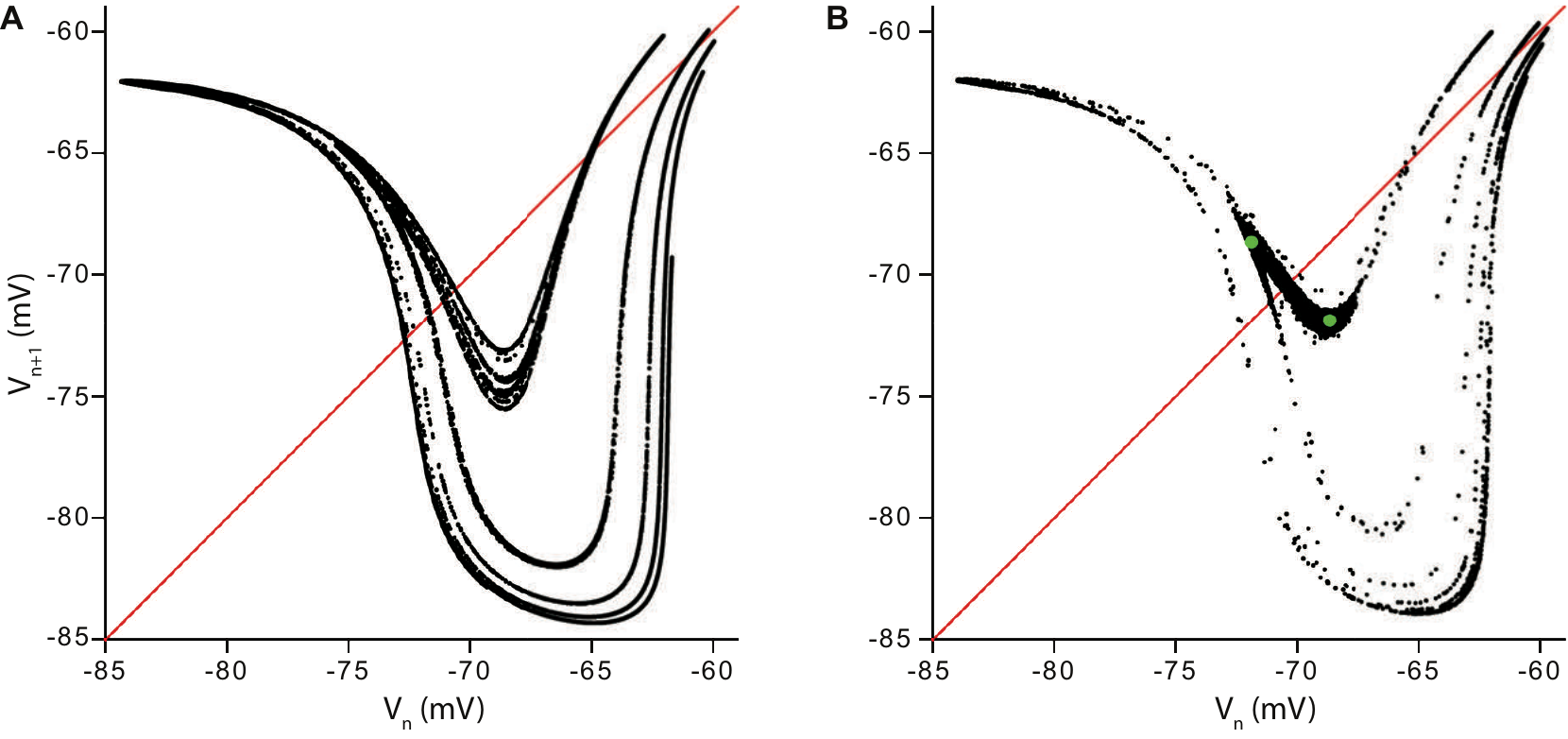}
	\caption{Period-doubling shown in  Poincar\'e return maps, $V_{\rm min}^{(n)} \to V_{\rm min}^{(n+1)}$,  with multiple unimodal branches. $\mathbf{A}$ Chaotic bursting in the deterministic model at $g_\mathrm{K1}=26.1226$ and $b=0.015$. $\mathbf{B}$ Weak ($\sigma=5\times 10^{-4}$) noise induces bursting that overshadows a period-2 obit (shown by two green dots) at  $g_\mathrm{K1}=25.972$ and $b=0.015$.}
	\label{fig:hairpd}
\end{figure} 
%

The resonance of a torus is determined by its winding number, calculated as the ratio of 2$\pi$ and the angle   $\phi$ of the Floquet multipliers $e^{\pm i \phi}$ on a unit circle \cite{Shilnikov2001ts}. For example, when the angle is $\pi/2$ (Fig.~\ref{fig:alien}$\mathbf{A_2}$), the winding number is 4, meaning that the resonance of the torus is 1:4. The irrational and rational winding numbers correspond to ergodic (quasi-periodic) and resonant (periodic) torus, respectively. The cases 1:1, 1:2, 1:3, 1:4 are considered as strong resonances and others as weak. Figure~\ref{fig:alien} shows the emergence, evolution and breakdown of the 2D tori through three strong resonances, namely 1:2, 1:3, 1:4, and one weak case.  When the torus becomes resonant, the corresponding IC (grey color) of the  Poincar\'e  return map possesses stable periodic orbits (shown as blue dots). The resonant ICs becomes non-smooth and then starts breaking down when the leading multipliers of the resonant periodic orbits   become complex conjugate towards forthcoming PD bifurcations (shown as green and orange dots in the return maps) with the increase of the parameter $g_\mathrm{K1}$ at the indicated values of parameter $b$, so that the unstable sets of the saddle periodic orbits no longer converge monotonically but start spiraling onto the stable orbits on the torus (see Fig.~\ref{fig:hairtsprocess}{\bf D}). 

\begin{figure}[h!]
	\includegraphics[width=.7\columnwidth]{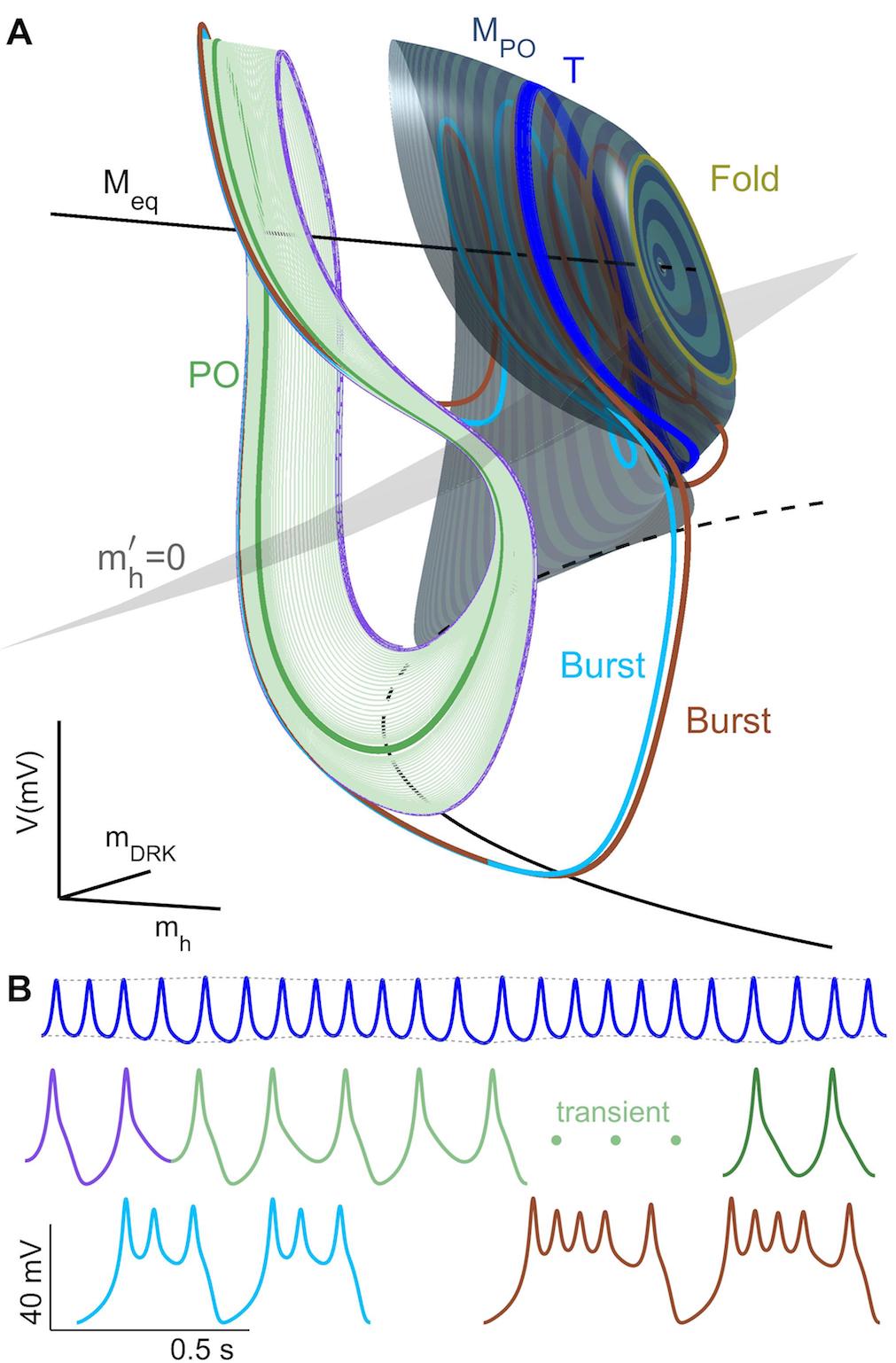}
	\caption{$\mathbf{A}$ 3D $\mathrm{(m_h, m_{DRK}, V)}$-phase space projection depicting a few exemplary tonic-spiking, quasi-periodic and bursting orbits being superimposed on the slow-motion manifolds: 1D S-shaped quiescent $\mathrm{M_{eq}}$ with two knees/folds, and the 2D tonic spiking $\mathrm{M_{PO}}$.  The fold on $\mathrm{M_{PO}}$ corresponds to a saddle-node bifurcation giving rise to periodic and  quasi-periodic tonic-spiking [torus as dark-blue orbits] occurring on the outward stable branch of $\mathrm{M_{PO}}$. The grey surface, $\mathrm{m^\prime_{h}=0}$, is the slow nullcline above/below which the slow variable increases/decreases during tonic-spiking/quiescent phases of bursting. $\mathbf{B}$ Voltage traces with matching colors of the corresponding orbits in $\mathbf{A}$.}
	\label{fig:hairnemo}
\end{figure}
%

\subsection{Period-doubling en route to bursting}
Next we consider the pathway at $b=0.015$ passing through the period-doubling curve (green line) in the 2D parameter plane of Fig.~\ref{fig:hair1}$\mathbf{A}$. Having crossed this curve, the stable periodic orbit (representing tonic-spiking oscillations) loses its stability to the one with the doubled period. This bifurcation and the new periodic orbit correspond to the branching in the diagram shown in Fig.~\ref{fig:hair2}$\mathbf{B}$. A cascade of forthcoming period-doubling bifurcations develops quite rapidly as $g_{\rm K1}$ increases, and terminates with the onset of regular bursting in the hair cell model. Spike adding causes the onset of chaotic bursting in the transition between stable 5-spike and 4-spike bursting as this bifurcation diagram reveals.

The Poincar\'e return map of the chaotic bursting shown in Fig.~\ref{fig:hairpd}{\bf A} has a wider drop-shaped section, compared to chaotic bursting emerging through the torus scenario (Fig.~\ref{fig:hairpoinc}{\bf B}). Quantitatively, the width of the drop indicates how fast the model switches back and forth between tonic spiking and hyperpolarized quiescence in bursting. The wider the drop in the map is, the less rigid the switching is, or, alternatively, the less contrast between fast and slow dynamics is. Since the switches are quite loose, not constrained, there is a wide variability in transitions that give rise to multiple branches of the return map in Fig.~\ref{fig:hairpd}{\bf A}.
The multiple branches look like as they represent snapshots of some return maps at several parameter values in progression. Each branch reveals a generic unimodal (non-sharp) shape of the return map that is needed for a cascade of period-doubling bifurcations to occur. We note that other Hodgkin-Huxley type models with the fold on the tonic spiking manifold can exhibit a period-doubling cascade leading to a chaotic tonic-spiking attractor in the phase space \cite{shilnikov2012complete}. The progression of the branches also suggests that the stable periodic orbit emerges through a saddle-node bifurcation when the graph of the map touches the $45^\circ$-degree line. Recall that the corresponding bifurcation comes out of the codimension-two Bautin point in the parameter plane in Fig.~\ref{fig:hair1}{\bf A}. Next, as the map graph lowers further and becomes steeper, the stable fixed point becomes unstable thus giving rise to a period-two orbit, as such indicated by two green solid circles in Fig.~\ref{fig:hairpd}$\mathbf{B}$. 
The bottom four branches of the map correspond to bursting where each hyperpolarized voltage drop is followed by chaotic tonic-spiking phases developed through all steps of the period doubling cascade.

Precursors of bursting can be revealed by adding small random perturbations to the model. Gaussian white noise $\xi(t)$ is injected to the right hand side of Eq.(\ref{eq:hair}) with the term  $\sigma \xi(t)$, where $\sigma$ is the standard deviation  of random perturbation. 
Fig.~\ref{fig:hairpd}$\mathbf{B}$ demonstrates that weak random perturbations can effectively reveal precursors of transition to bursting. 
For the value of $g_\mathrm{K1}$, the model possesses a stable period-2 orbit and so the Poincar\'e return map contains two points only.
Weak noise induces bursting resulting in the map which has a similar shape as the deterministic chaotic bursting.

We stop here to resume that without 1D Poincar\'e return maps generated through voltage oscillations, it would be impossible to characterize the exact formation mechanisms underlying transitions from simple tonic spiking to bursting and back in this model as well as in other Hodgkin-Huxley type
  neuronal models \cite{Channell2007a,Channell2009}.        

\begin{figure}[h!]
	\includegraphics[width=\columnwidth]{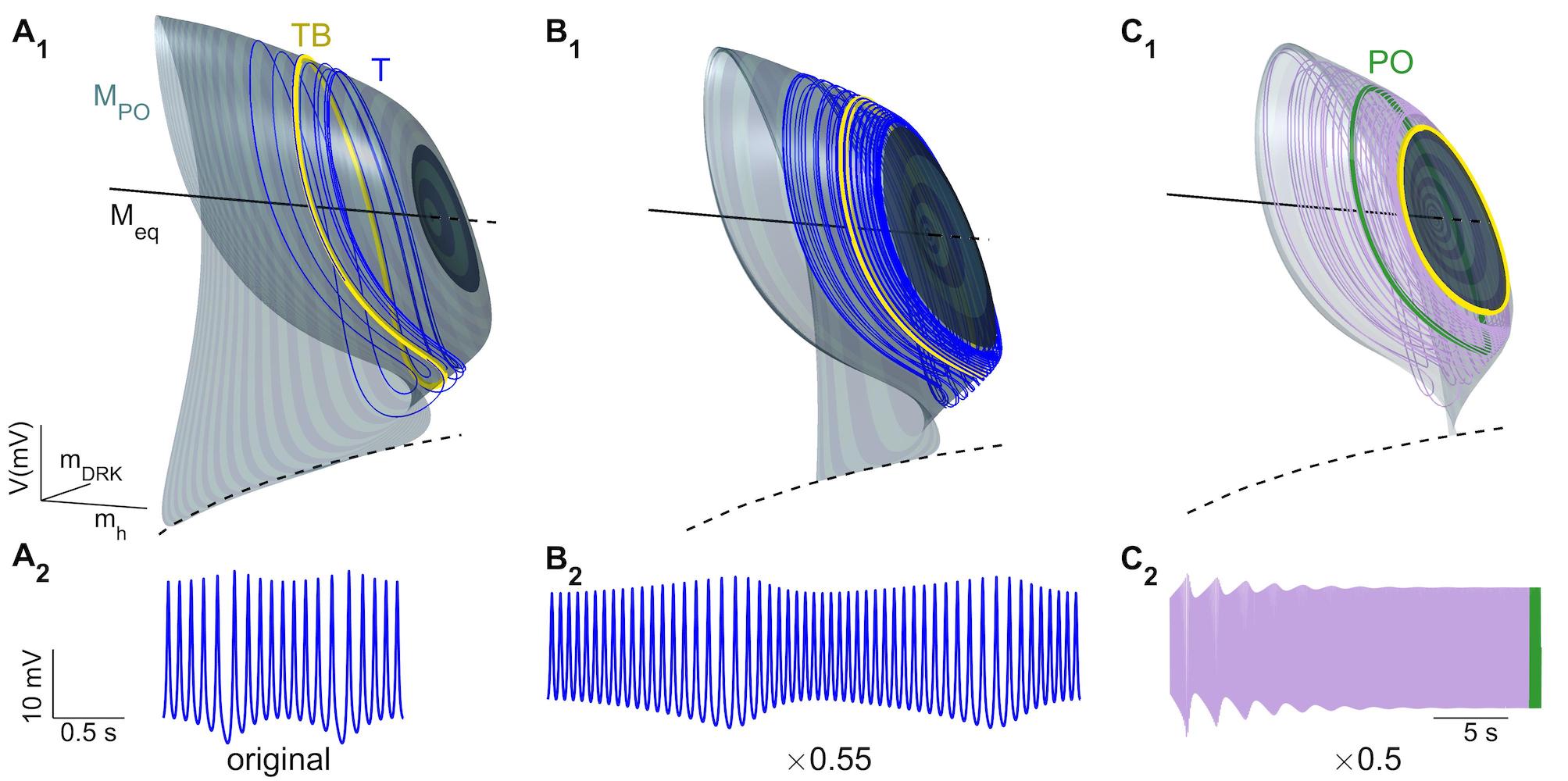}
	\caption{Scaling down the slow $\mathrm{m_{h}}$-variable shifts the emergent torus closer to the fold and  changes its stability in the hair cell model. $\mathbf{A_{1}}$--$\mathbf{B_{1}}$ 3D $\mathrm{(m_h, m_{DRK}, V)}$-phase space projection show the stable torus (blue line) emerging from a PO (yellow ring, detected by MATCONT) on stable section of $\mathrm{M_{PO}}$, for the original and 0.55 slowed rate of $\mathrm{m_{h}}$-gating variable, resp. $\mathbf{C_{1}}$: Further decreasing the rate by the factor of two makes the unstable torus (of the saddle type) emerging at the fold on $\mathrm{M_{PO}}$. The saddle torus creates bistability as bounds the stable tonic orbit (green circle) away from bursting (see Fig.~\ref{fig:hairbi}). Insets $\mathbf{A_{2}}$--$\mathbf{C_{2}}$ depict the corresponding voltage traces of the tonic-spiking attractors above.}
	\label{fig:hairpush}
\end{figure}
%

\subsection{Parameter continuation \& slow-fast dissection}
Bifurcations underlying the transitions between different oscillatory patterns in the hair cell model can be explained geometrically using the slow-fast dissection as illustrated in Fig.~\ref{fig:hairnemo}. 

The essence of the approach is based on dissecting the original dynamics of the model into slow and fast components. In the model the activation of the $h$-current, $m_{ \rm h}$, is the slowest dynamical variable, while all other variables are fast compared to it.  
In particular, we choose the membrane potential, $V$, and the activation of the K$^+$ DRK-current, $m_{\rm  DRK}$, as representatives of the fast subsystem. The feature of a slow-fast system that often reduces the complexity of its dynamics is that the phase trajectory stays close to the  slow-motion manifolds of low dimensions: a 2D tonic-spiking manifold, $\mathrm{M_{PO}}$, foliated by small-amplitude periodic orbits,  and a 1D $S$-shaped quiescent manifold, $\mathrm{M_{eq}}$, comprised of hyper- and depolarized equilibrium states of the model. This approach implies that bursting, being a multiple time-scale process, is interpreted as repetitive fast switching between these manifolds at their terminal phases, followed by slow passages of trajectories turning around $\mathrm{M_{PO}}$ and then along $\mathrm{M_{eq}}$, corresponding to the episodes of tonic-spiking and hyperpolarized quiescence, respectively. Figure \ref{fig:hairnemo} depicts the topology of these slow-motion manifolds for the parameters close to the torus (dark blue trajectory) bifurcation, as well as shows several bursting orbits superimposed (green, light blue and brown trajectories) in the phase-space projection. These spiking and quiescent manifolds are obtained by the parameter continuation of periodic orbits and equilibria of the model with the auxiliary parameter shifting the slow nullcline $m^\prime_{ \rm h}=0$ up and down.  

\begin{figure}[t!]
	\includegraphics[width=.7\columnwidth]{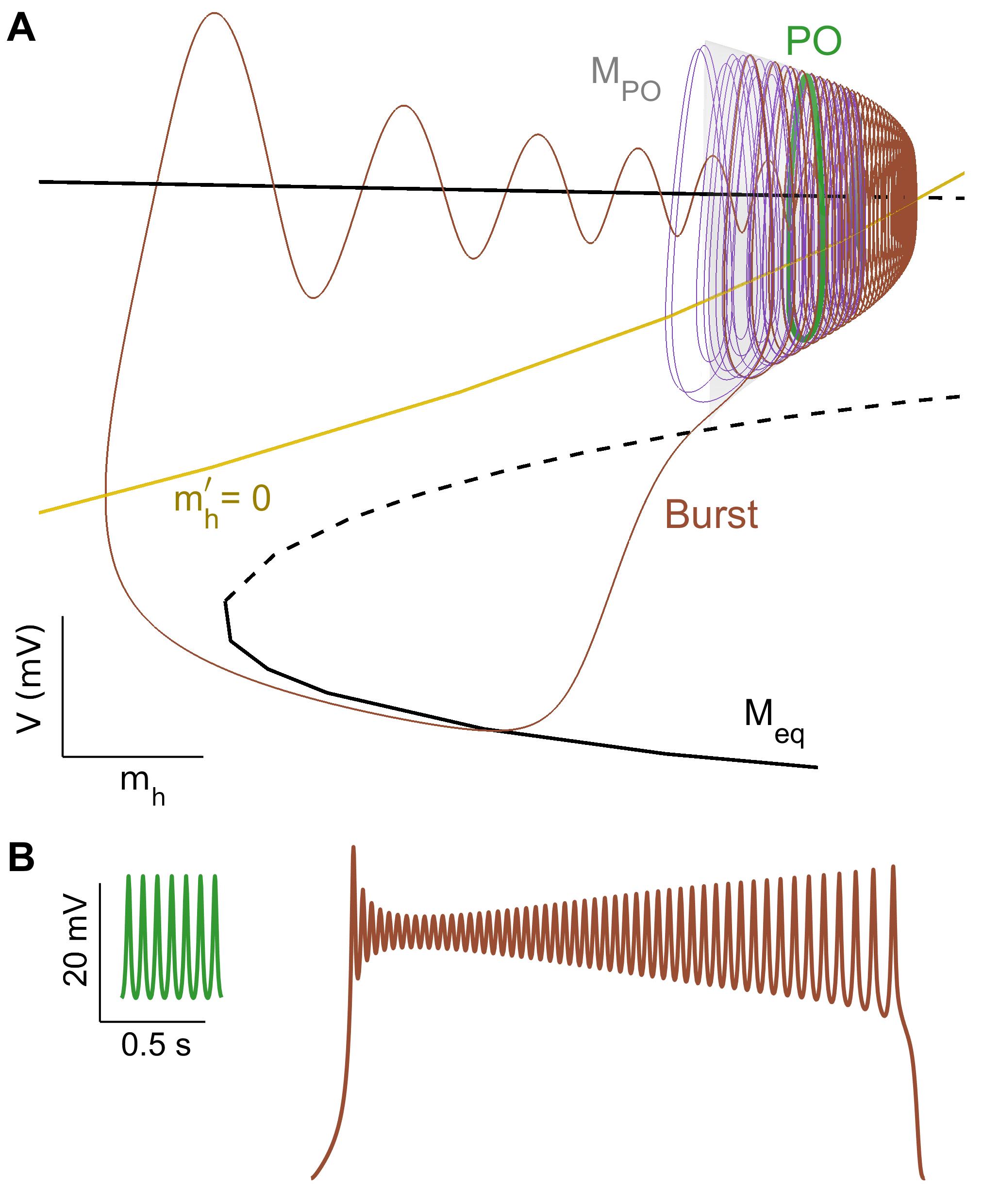}
	\caption{$\mathbf{A}$ 2D $\mathrm{(m_h, V)}$-phase space projection shows bistability of tonic spiking orbit (green circle, PO) and bursting orbit (brown) being overlaid on the quiescent manifold, $\mathrm{M_{eq}}$, and the tonic-spiking manifold, $\mathrm{M_{PO}}$, when the $\mathrm{m_{h}}$-gating variable of the hair cell model is scaled down two times slower than its original rate. These two stable states are separated by a saddle torus, denoted here by the transient part of the tonic spiking orbit (pink). Inset $\mathbf{B}$ shows the  voltage traces corresponding to the tonic spiking and bursting orbits in $\mathbf{A}$.}
	\label{fig:hairbi}
\end{figure}

In what follows we outline how the manifolds are detected in the 12D phase space of the hair cell model. We use a new and practical way for localization of both slow motion manifolds, 1D quiescent, $\mathrm{M_{eq}}$, and 2D tonic-spiking, $\mathrm{M_{PO}}$, composed of equilibria and periodic orbits of the full system. The method capitalizes on the slow-fast dissection as well as on the parameter continuation technique \cite{dhooge2003matcont}. The methods supplements the conventional slow-fast dissection in the singular limit with  the slow variable frozen as a parameter.  The main essence is the parameter continuation technique applied to the entire set of the model equations. The advantage of our approach is that it yields the sought manifolds themselves in the phase space of the intact model rather the manifolds of its fast subsystem without taking into account slow yet finite component of the overall dynamics. This approach is especially valuable for complex models of higher dimensions where slow–fast dissections could be problematic because of multiple time-scales various ionic currents involved in complex dynamics especially at transitions. Furthermore, this approach can detect and explains bifurcations, often non-local due to reciprocal interactions of slow and fast  dynamical variables that underlie these transitions in the model in question.    

Specifically, we introduce a faux control parameter, which is a voltage offset that affects only the slowest dynamics of the model; this offset shifts the voltage threshold at which the slow  gating variable/channel is half-open.  Geometrically, its variations lift or lower the slow nullcline, $\mathrm{m^\prime_h=0}$, in the 3D phase-space projection depicted in Fig.~\ref{fig:hairnemo}. As the voltage offset changes, the intersection point of the slow nullcline (sigmoidal surface) with $\mathrm{M_{eq}}$ [the equilibrium state of the model] is shifted along $\mathrm{M_{eq}}$, thus tracing $\mathrm{M_{eq}}$ down and explicitly revealing its position and shape in the phase space. Same is true for the other manifold $\mathrm{M_{PO}}$. More details on this approach can be found in \cite{shilnikov2012complete}.

Like many neuron models, the quiescent manifold in the hair cell system has a characteristic $S$-shape in the projection onto the slow-fast ($\mathrm{m_{h}},\mathrm{V}$)-plane. The shape of $\mathrm{M_{eq}}$ implicitly endows the model with a hysteresis required for dynamic switching between hyper- and depolarized phases in large-amplitude bursting. Reaching the low, hyperpolarized fold on $\mathrm{M_{eq}}$ indicates the beginning of a burst. In addition, the two real bifurcation parameters, $g_{\rm  K1}$ and $b$, of the model can shift the position of the manifold $\mathrm{M_{eq}}$ relative to that of the slow nullcline, $\mathrm{m^\prime_h=0}$ in the phase space, so that the intersection point can move onto the stable upper or low section of $\mathrm{M_{eq}}$. In those cases, the  hair cell rests on the depolarized or hyperpolarized steady state, respectively.  
\begin{figure}[t!]
	\includegraphics[width=.7\columnwidth]{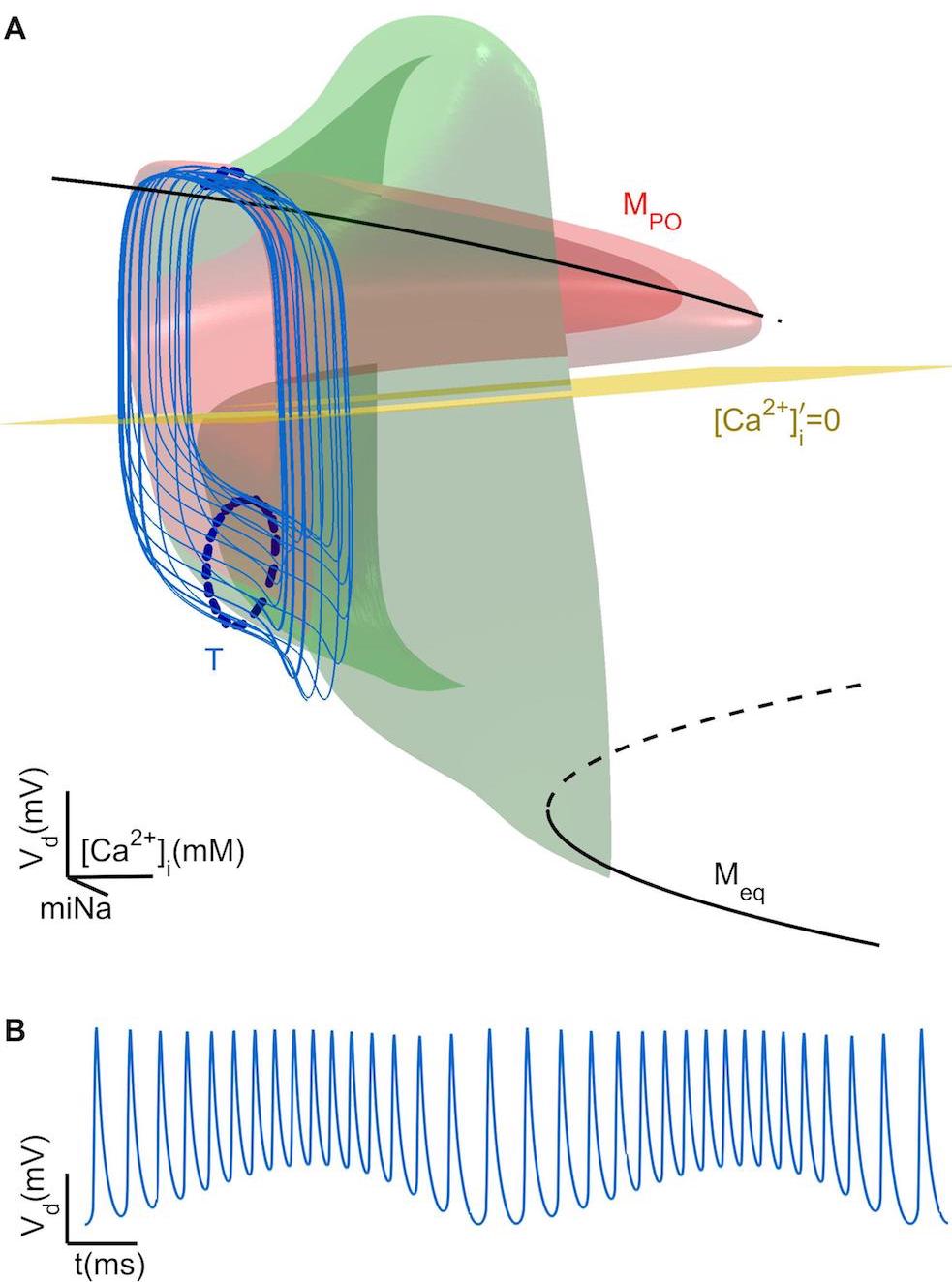}
	\caption{$\mathbf{A}$ 3D $\mathrm{([Ca^{2+}]_{i}, m_{iNa}, V)}$-phase space projection of the pyramidal cell model shows the stable torus (light blue line) on the fold of slow-motion manifolds, $\mathrm{M_{PO}}$, superimposed on the 1D S-shaped quiescent manifold, $\mathrm{M_{eq}}$, and the slow nullcline, $\mathrm{[Ca^{2+}]_{i}^\prime=0}$ (yellow surface). The Poincar\'{e} cross-section transversal to the torus highlights the dark blue dots constituting the stable IC on it. Inset $\mathbf{B}$ shows the slowly modulated voltage trace corresponding to the torus in $\mathbf{A}$.}  
	\label{fig:pyramid}
\end{figure}

The hair cell model oscillates tonically when there is a stable periodic orbit on $\mathrm{M_{PO}}$. This orbit emerges from the depolarized equilibrium state through the supercritical AH bifurcation (discussed above). Moving the slow nullcline by varying the faux parameter makes the periodic orbit slides along $\mathrm{M_{PO}}$, thus revealing its shape. The manifold ends up through a homoclinic bifurcation occurring after the periodic orbit of the increased size and touches the middle, saddle branch of $\mathrm{M_{eq}}$ to become the homoclinic obit of the saddle. This is a generic configuration for square-wave bursters.   
\section{Purkinje cell model}
%
\begin{figure}[t!]
	\includegraphics[width=0.9\columnwidth]{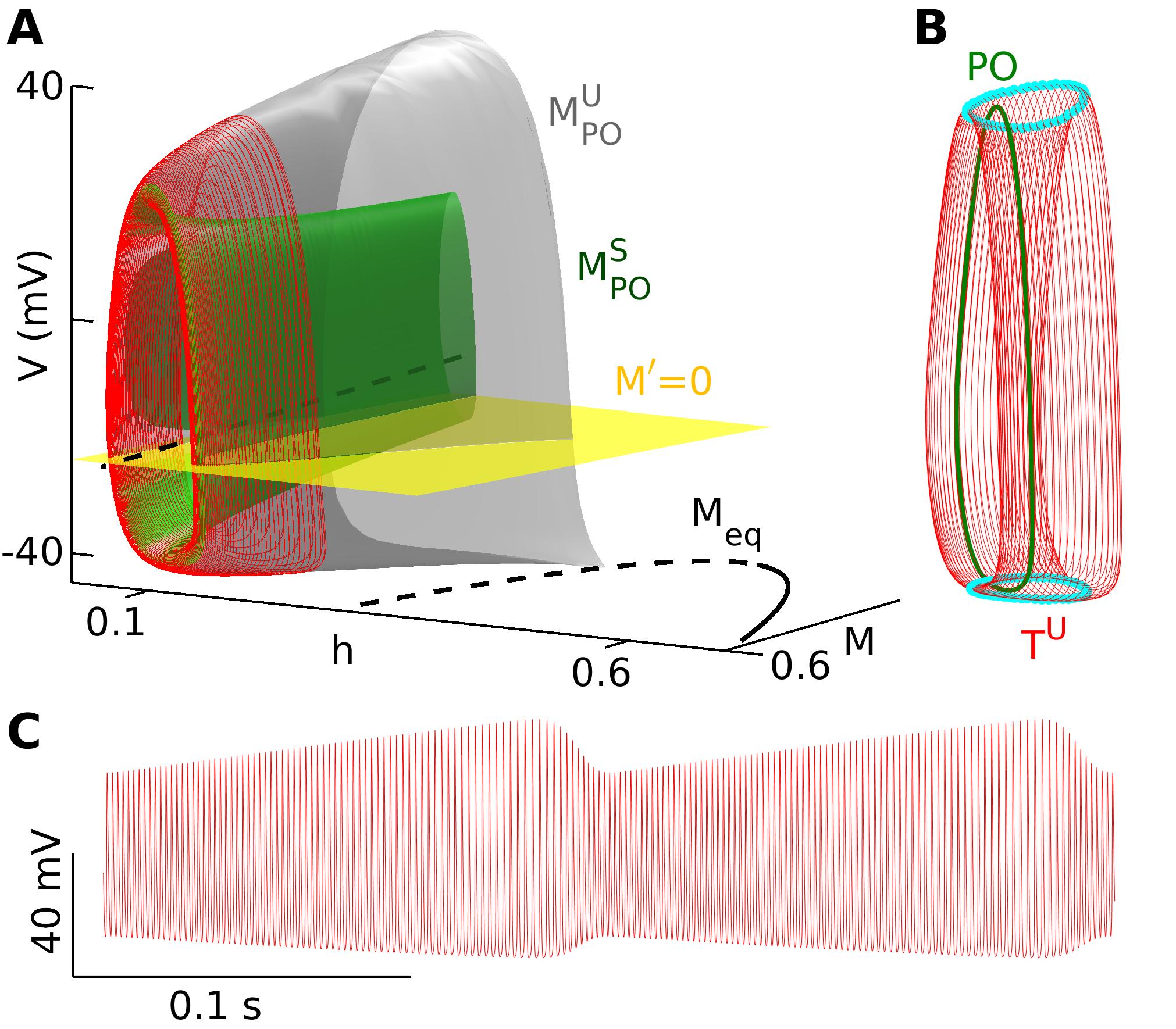}
	\caption{\textbf{A} 3D $\mathrm{(h, M, V)}$-phase space projection of the  Purkinje cell model at $\mathrm{I_{app}}=-29.487$ depicting the saddle (red) torus-canard slowly oscillating between the stable (green) inward $\mathrm{M_{PO}^S}$ and unstable (grey) outward $\mathrm{M_{PO}^U}$ branches merging at the fold of the 2D spiking manifold. The yellow surface is the slow nullcline $\mathrm{M'=0}$. The 1D quiescent manifold $\mathrm{M_{eq}}$ (black line) has stable (solid) and unstable (dashed) branches. \textbf{B} Magnified saddle torus-canard ${\mathrm {T^U}}$ (red) encloses a stable tonic-spiking PO (green circle). The phase points at maximum and minimum voltage values trace out two circles on the space space. \textbf{C} Slowly modulated voltage trace in \textbf{A}.}
	\label{fig:torusNemo}
\end{figure}

\begin{figure}[h!]
	\includegraphics[width=.9\columnwidth]{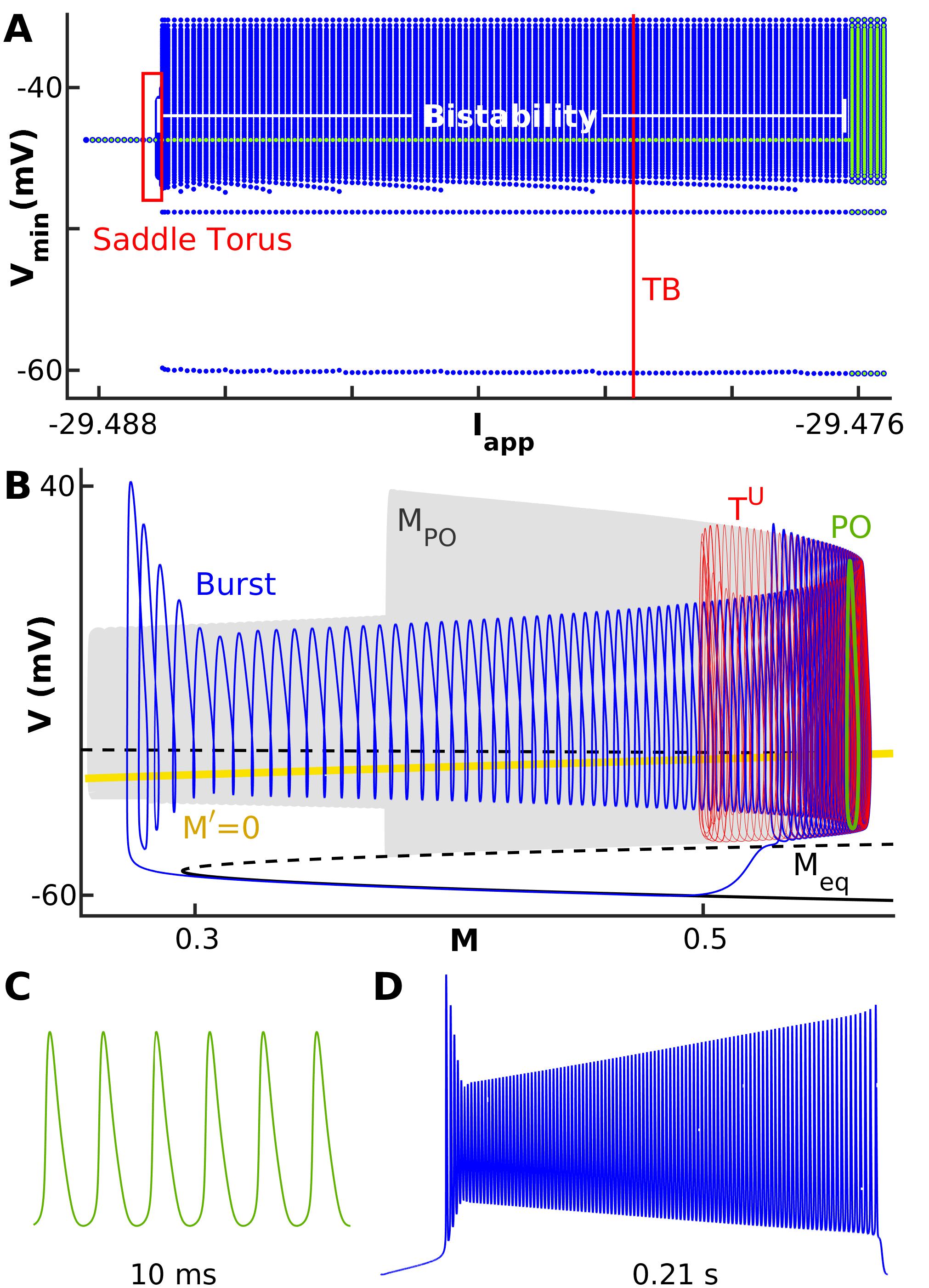}
	\caption{Bistability of tonic-spiking and bursting in the Purkinje cell model.
		$\mathbf{A}$ Bifurcation diagram represented by the bi-directional forward/backward $\mathrm{I_{app}}$-sweeps of $\mathrm{V_{min}}$-values (green/blue dots, resp., and red box for saddle torus-canard). White horizontal bar indicates the range of the co-existence of stable tonic-spiking and bursting, red vertical line at $\mathrm{I_{app}=-29.4796}$ is where MATCONT detects the TB. $\mathbf{B}$ 2D $\mathrm{(V, M)}$-phase projection depicting the co-existing stable tonic-spiking (green) and bursting (blue) orbits separated by the saddle torus-canard (red) orbit superimposed on the manifold $\mathrm{M_{PO}}$ (grey), the slow nullcline $\mathrm{M'=0}$ (yellow), and the manifold $\mathrm{M_{eq}}$ (black line) at $\mathrm{I_{app}=-29.487}$. Insets  $\mathbf{C}$ and $\mathbf{D}$ show the voltage traces corresponding to the tonic-spiking and bursting orbits in $\mathbf{B}$, resp.}  
	\label{fig:purkBi}
\end{figure}

Whenever neither the stable branch of $\mathrm{M_{PO}}$ nor the stable branches of the quiescent manifold $\mathrm{M_{eq}}$ are cut through by the slow nullcline, $\mathrm{m^\prime_h=0}$, i.e., both manifolds are freely transient, the hair cell model exhibits bursting. Bursting is represented by solutions repeatedly switching between the low, hyperpolarized branch of $\mathrm{M_{eq}}$ and the tonic spiking manifold $\mathrm{M_{PO}}$. The number of complete revolutions of the solution around $\mathrm{M_{PO}}$ is that of spikes per burst in the voltage traces. This winding number is used to classify the bursting activity. The larger the number is, the longer the burst duration lasts.
The relative position of the torus to the fold on $\mathrm{M_{PO}}$ can be changed by the rate of the slow variable $\mathrm{m_{h}}$. The slower the rate, the closer the stable torus emerges to the fold as shown in Figs.~\ref{fig:hairpush}$\mathbf{A}$--$\mathbf{B}$, as well as the closer the torus bifurcation (TB) is detected near the fold by the bifurcation package MATCONT \cite{content, dhooge2003matcont}.   More interestingly, when the rate of $\mathrm{m_{h}}$ is about two times slower, the torus bifurcation becomes sub-critical and generates a saddle torus at the fold instead (see Fig.~\ref{fig:hairpush}$\mathbf{C}$). This gives rise to bistability in the system, where both tonic spiking and bursting can be produced by the model depending on the initial conditions, as illustrated  n  Fig.~\ref{fig:hairbi}. A small basin of attraction of the stable tonic-spiking orbit is bounded by the 2D saddle torus in the phase space from that of the stable bursting orbit. We note though that similar bistability is also found in the Purkinje cell model, parabolic burster model and FitzHugh-Nagumo-Rinzel (FNR) model (see the following sections).  

In a system with drastically distinct slow and fast time scales, the periodic orbit would bifurcate into a torus or through period-doubling right at the location of the fold \cite{Wojcik2011}. The type of the bifurcation depends on global structure of the flow at the fold. Specifically, depending on whether the phase space volume  contracts or not,  
the tonic spiking orbit undergoes period-doubling or torus bifurcation. While the verification of this condition seems doable in 3D slow-fast systems (see the FNR section below),
the high-order models may present a challenge as one has to identify the set of the variables (from a 3D subspace containing the central manifold in which the bifurcation takes place) that are involved directly into the ongoing bifurcation.

\section{Torus in a Pyramidal Cell Model}

Our next example is another highly detailed 14D Hodgkin-Huxley type two-compartmental (for soma and dendrite) model  describing interactions of pyramidal cells and fast-spiking inhibitory interneurons \cite{mainen1996influence}. The reader can find details of all currents constituting the model  in Ref. \cite{krishnan2015electrogenic}, which describes how electrogenic properties of the ${\mathrm Na^+}/{\mathrm K^+}$ ATPase control transitions between normal and pathological brain states, specifically during epileptic seizures.  The equations describing the dendritic and somatic voltage dynamics are:  
\begin{eqnarray} 
&&C_\mathrm{m} V_{\mathrm{d}}'= -I_\mathrm{d}^{\mathrm{int}}-I_\mathrm{d}^{\mathrm{leak}}-\alpha I_\mathrm{d}^{\mathrm{pump}}-\frac{g_\mathrm{c}(V_\mathrm{d}-V_\mathrm{s})}{s_\mathrm{d}}, \nonumber\\
&&\frac{g_\mathrm{c}(V_\mathrm{d}-V_\mathrm{s})}{s_\mathrm{d}}=-I_\mathrm{s}^{\mathrm{int}}-I_\mathrm{s}^{\mathrm{leak}}-\alpha I_\mathrm{s}^{\mathrm{pump}} \label{pyramideq}, 
\end{eqnarray}
where $C_{m}$ is the capacitance of the dendritic compartment, $V$ is the  membrane potential, $I$s are various  ionic currents including $\mathrm{Na^+}$/$\mathrm{K^+}$ pump current, \textit{s} stands for a surface of each compartment, $g_{\mathrm{c}}$ is the coupling conductance of the soma (s) and the dendrite (d),  $\alpha$ is a scaling coefficient. 

Unlike in the hair cell model, where the dynamics of intracellular calcium concentration, $\mathrm{[Ca^{2+}]_{i}}$, is relatively fast, here 
$\mathrm{[Ca^{2+}]_{i}}$ is the slowest dynamical variable in the model. As for the hair cell model, we will use the 3D phase projection to show the topology of the slow-motion manifolds that determine the shape of bursting oscillations in the model. Figure~\ref{fig:pyramid} depicts a similarly shaped  1D quiescent manifold  $\mathrm{M_{eq}}$. Its lower, hyper-polarizing knee corresponds to the homoclinic saddle-node bifurcation giving rise to the stable (green)  outer section of the 2D tonic-spiking manifold $\mathrm{M_{PO}}$. This manifold wraps back inwardly so that its unstable (pink) shrinking section terminates on the depolarizing branch of $\mathrm{M_{eq}}$  through a  sub-critical AH bifurcation. The surface $\mathrm{[Ca^{2+}]_{i}^\prime=0}$ (yellow) is the slow nullcline: $\mathrm{[Ca^{2+}]_{i}}$ increases/decreases above/below it in the phase space. When this nullcline is slightly below its position depicted in Fig.~\ref{fig:pyramid}, there would be a stable periodic orbit on the stable branch of $\mathrm{M_{PO}}$ that corresponds to tonic-spiking activity. As the nullcline  $\mathrm{[Ca^{2+}]_{i}^\prime=0}$ is shifted up, the stable orbit comes close to the fold and looses its stability through a supercritical torus bifurcation. The stable newborn torus oscillates back and forth between the stable and unstable branches of $\mathrm{M_{PO}}$  near the fold. The torus is stable because it lingers longer on the stable branch then on the unstable one on average. Otherwise it would be of the saddle type, i.e. repelling in the $\mathrm{m_{Na}}$ and $\mathrm{[Ca^{2+}]_{i}}$ coordinates and stable in the all other dimensions. 
One can observe from Fig.~\ref{fig:pyramid}{\bf A} that gaps between the successive points filling in the stable invariant curve are large enough to assume that dynamics of $\mathrm{[Ca^{2+}]_{i}}$ is not that slow compared to the torus-canard observed Purkinje cell model discussed next. After the torus breaks down as the slow nullcline is further shifted up, the pyramidal cell model demonstrates bursting activity, see ~\cite{krishnan2015electrogenic}.

\begin{figure}[t!]
	\includegraphics[width=\columnwidth]{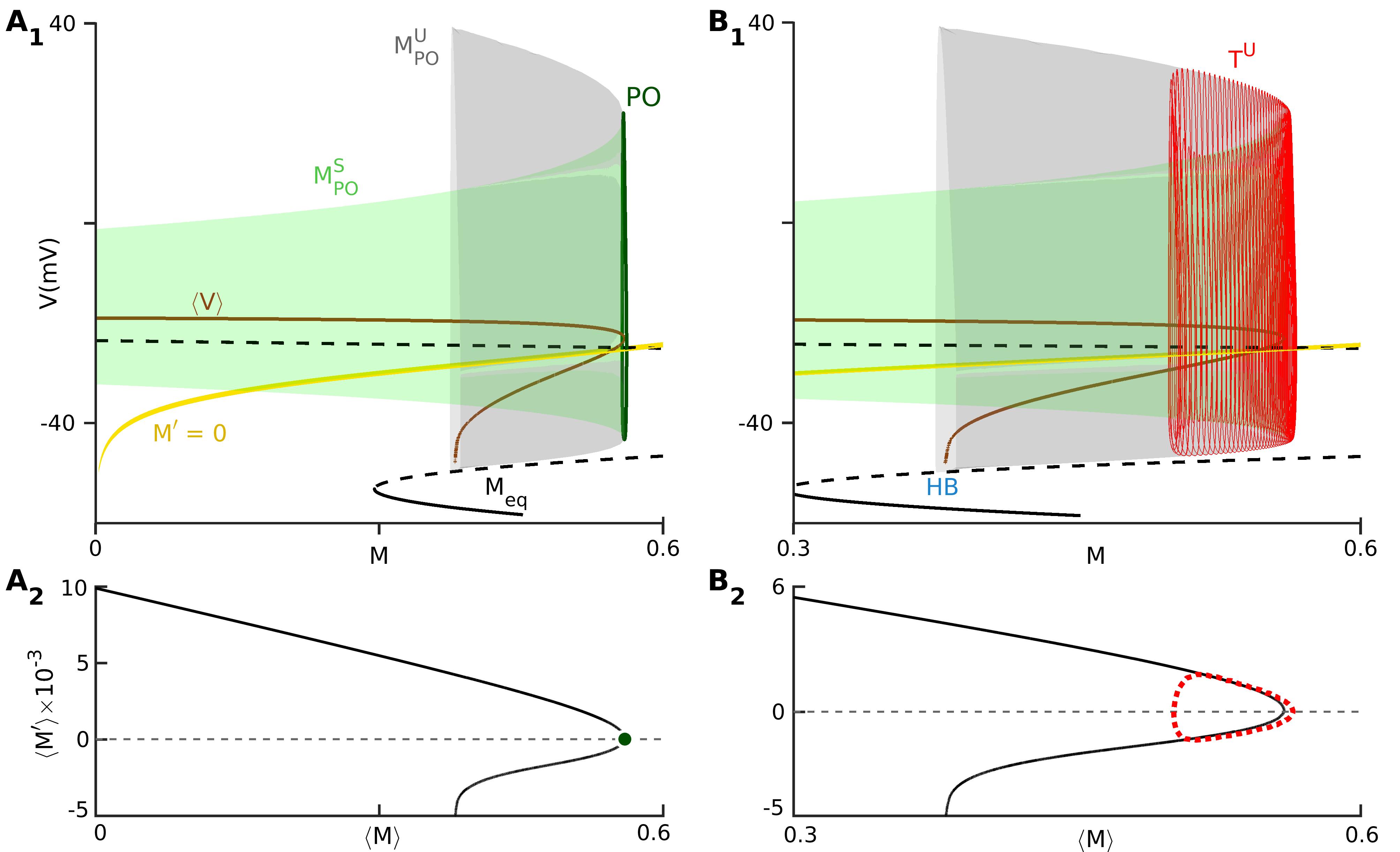}
	\caption{Averaging technique for the Purkinje cell model. $\mathbf{A_1}$--$\mathbf{B_1}$ $\mathrm{(V, M)}$-projections of the stable $\mathrm{M_{PO}^S}$ (green) and unstable $\mathrm{M_{PO}^U}$ (grey) sections of the 2D tonic-spiking manifold, the 1D S-shaped quiescent manifold $\mathrm{M_{eq}}$ (black line), slow nullcline $\mathrm{M'=0}$ (yellow line) and the 1D curve $\mathrm{\langle V \rangle}$ (brown) made up of the averaged phase coordinates of the POs constituting $\mathrm{M_{PO}}$. $\mathbf{A_1}$ Basin of the stable tonic-spiking PO (dark green) is bounded by the saddle torus, shown in $\mathbf{B_1}$ (red), at $\mathrm{I_{app}=-29.487}$. Widening unstable (grey) branch $\mathrm{M_{PO}^U}$ bending outward from the fold  terminates on the middle, saddle branch of $\mathrm{M_{eq}}$ after the homoclinic bifurcation (HB). Insets $\mathbf{A_2}$--$\mathbf{B_2}$ disclose the graph (black line) of the average function $\mathrm{\langle M' \rangle}$  determining the dynamics of the slow $\mathrm{M}$-variable on $\mathrm{M_{PO}}$;  a single zero (dark green dot) of $\mathrm{\langle M' \rangle}$ and its slope determine the position and the stability of the corresponding PO in the $\mathrm{M}$-direction. Inset $\mathbf{B_2}$ shows the stable IC (made up of red dots), corresponding to the 2D torus in $\mathbf{B_1}$, oscillating around the fold on the graph of $\mathrm{\langle M' \rangle}$. }  
	\label{fig:average}
\end{figure}

Purkinje cells are the main output of the cerebellum and are involved in motor learning/conditioning\cite{kandel2000principles}.  To study these cells, several highly detailed models were constructed  based on experimental data\cite{miyasho2001low,middleton2008high}. A reduced yet quantitatively-similar 5D Hodgkin-Huxley type model was proposed and studied in \cite{Kramer2008a}. Its generic representation reads as follows:  
\begin{align}
\nonumber \mathit{V}' = &  - I_{\rm app}-g_{K}\mathit{n}^4 \left(\mathit{V}+95\right)-g_{Na}{m_{0}}^3h\left(\mathit{V}-50\right)\\
\nonumber & -g_{L}\left(V+70\right)-g_{Ca}\mathit{c}^2\left(V-125\right) \\
\nonumber & -g_{M}\mathit{M}\left(V+95\right),\\
{\mathbf x}' = & \left(f_{\infty}\left[V+\Delta\right]-{\mathbf x}\right)/\tau_{{\mathbf x}}\left[V\right],
\label{eq:purkinje}
\end{align}
where the vector ${\mathbf x}$ represents the gating variаbles: fast \textit{n}, \textit{h}, \textit{c} and slow \textit{M}, for the following four ionic currents: a delayed rectifier potassium current, a transient inactivating sodium current, a high-threshold non-inactivating calcium current and a muscarinic receptor suppressed potassium current; here the external current $I_{\rm app}$ is the bifurcation parameter of the model. The Reader is welcome to consult with the original paper \cite{Kramer2008a} for more detailed description of the model and Appendix for its equations and Matlab code.

In this 5D model, the gating variable $\mathit{M}$ is the slowest one. A faux parameter, $\Delta$, is introduced in the right-hand side of the equation (see Eqs.~(\ref{eq:purkinje})) governing $\mathit{M}^\prime$  to perform the parameter continuation letting us sweep the slow-motion manifolds using the package  MATCONT \cite{dhooge2003matcont} without affecting the fast subsystem of the model .  

Figure~\ref{fig:torusNemo}$\mathbf{A}$ shows that the tonic-spiking manifold with the distinct fold consists of two components: the inner stable (green) cyllinder-shaped M$_{\mathrm{PO}}^{\mathrm{S}}$ and the outer unstable (red) part M$_{\mathrm{PO}}^{\mathrm{U}}$. Note that M$_{\mathrm{PO}}^{\mathrm{U}}$ vanishes via the homoclinic bifurcation (HB) when it touches the middle, saddle branch of the quiescent manifold M$_{\mathrm{eq}}$ ($S$-shaped black line); see also 2D phase-space projections in Figs.~\ref{fig:purkBi} and \ref{fig:average}).\\
It was first reported in the original paper\cite{Kramer2008a} that during the transition from tonic spiking to bursting, the Purkinje cell model shows slow amplitude-modulation of tonic-spiking activity in the voltage traces   (Fig.~\ref{fig:torusNemo}$\mathbf{C}$).  This modulation corresponds to a torus-canard that emerges around the fold of the tonic spiking manifold M$_{\mathrm{PO}}$ in the phase space of the model (see Fig.~\ref{fig:torusNemo}$\mathbf{A}$ and  $\mathbf{B}$). This torus-canard exists in a rather narrow parameter interval beyond which the model quickly transforms into a square-wave buster.   

\subsection{Saddle torus-canard and bistability}
To examine the stability of the torus in question, we tracked the stable state of the model by varying the applied current $\mathrm{I_{app}}$ in small incremental steps ($\sim 1\times10^{-4}$) in both directions - increasing and decreasing.  Simulations for each applied current value were long enough for the model to reach the steady state. We then represented the stable sates by minima of the voltage, shown in Fig.~\ref{fig:purkBi}$\mathbf{A}$. \\
Interestingly, for $\mathrm{I_{app}\in[-29.4871, -29.4762]}$ (denoted by white bar in Fig.~\ref{fig:purkBi}$\mathbf{A}$), the stable state in 
$\mathrm{I_{app}}$-increasing direction is tonic spiking (green dots in Fig.~\ref{fig:purkBi}$\mathbf{A}$), while in $\mathrm{I_{app}}$-decreasing direction is bursting (blue dots). That is, the system is bistable within that small window of $\mathrm{I_{app}}$. What separates the two stable states is a saddle torus-canard detected at $\mathrm{I_{app}=-29.487}$ (red box), whose corresponding voltage trace shows slow amplitude-modulation (Fig.~\ref{fig:torusNemo}$\mathbf{C}$). The phase-space trajectories corresponding to the two stable states and the separating torus-canard are shown in Fig.~\ref{fig:purkBi}$\mathbf{B}$ and the corresponding voltage traces are shown in Fig.~\ref{fig:purkBi}$\mathbf{C}$ and $\mathbf{D}$. The vertical (red) line indicated the I$_{\rm app}$ where MATCONT detects torus bifurcation.  

\begin{figure}[t!]
	\includegraphics[width=.8\columnwidth]{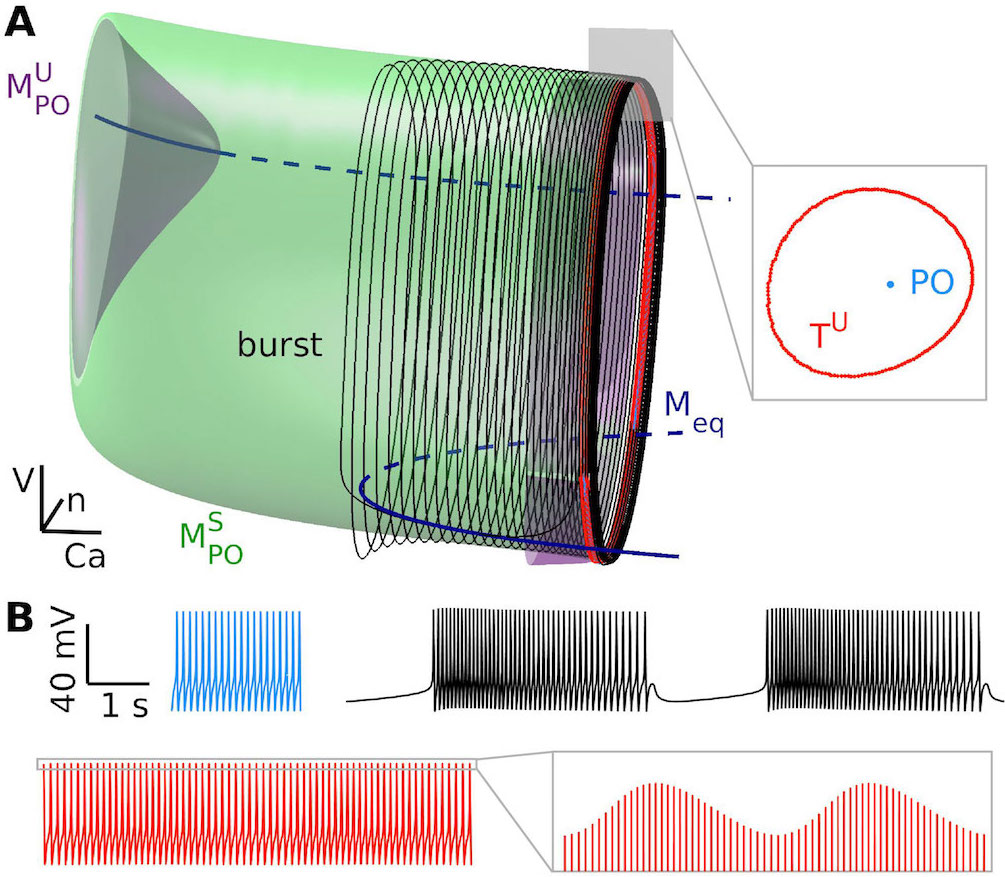}
	\caption{3D $(\mathrm{Ca,n,V})$-phase-space projection of the parabolic burster model. $\mathbf{A}$ Saddle torus (red line), enclosing a stable tonic-spiking (blue line) and a stable bursting orbit (black line) orbits are superimposed on the 1D quiescent manifold, $\mathrm{M_{eq}}$ (dark blue line), and the 2D tonic-spiking manifold with stable $\mathrm{M_{PO}^S}$ (green surface) and unstable $\mathrm{M_{PO}^U}$ (purple surface) sections merging at the fold. Magnified inset depicts  a local Poincar\'{e} cross-section  (grey plane at $n = 0.22$) with a repelling (red) IC circle, $\mathrm{T^{U}}$, enclosing the basin of the stable FP (light blue dot) corresponding to the tonic-spiking PO. Inset $\mathbf{B}$ shows the corresponding voltage traces of the  orbits in $\mathbf{A}$; parameters are given in Appendix.}  
	\label{fig:plant}
\end{figure}

\subsection{Averaging technique}
As for most slow-fast systems, the location of the tonic-spiking periodic orbit on the manifold in the phase space in the Purkinje cell model at various parameters can be accurately predicted by 
the averaging technique, see \cite{shilnikov2012complete} for details. The approach is illustrated in Fig.~\ref{fig:average}$\mathbf{A}$. Its core is the use of the parameter continuation to detect  periodic orbit(s) on M$_{\mathrm{PO}}$ where the average derivative of the slow variable vanishes as well as to find the corresponding function in the right-hand side of the average slow equation to predict forthcoming bifutcations like the blue sky catastrophe \cite{shilnikov2012complete}.  Specifically for the Purkinje cell model, first, $\mathrm{\langle M' \rangle}$ and $\mathrm{\langle M \rangle}$ are calculated for each periodic orbit on the tonic-spiking manifold based on the continuation data; then $\mathrm{\langle M' \rangle}$, is plotted against $\mathrm{\langle M \rangle}$ (which represents each periodic orbit on the tonic spiking manifold)  to pin down where the stable tonic spiking occurs (Fig.~\ref{fig:average}$\mathbf{A_2}$). The prediction from the averaging technique showing where $\mathrm{\langle M' \rangle=0}$ (dark green dot in Fig.~\ref{fig:average}$\mathbf{A_2}$) is quite precise, as it matches very well the result of the parameter continuation  simulations (dark green PO in Fig.~\ref{fig:average}$\mathbf{A_1}$). 
%

\section{FitzHugh-Nagumo-Rinzel model} 
\begin{figure*}[t!]
	\includegraphics[width=.85\textwidth]{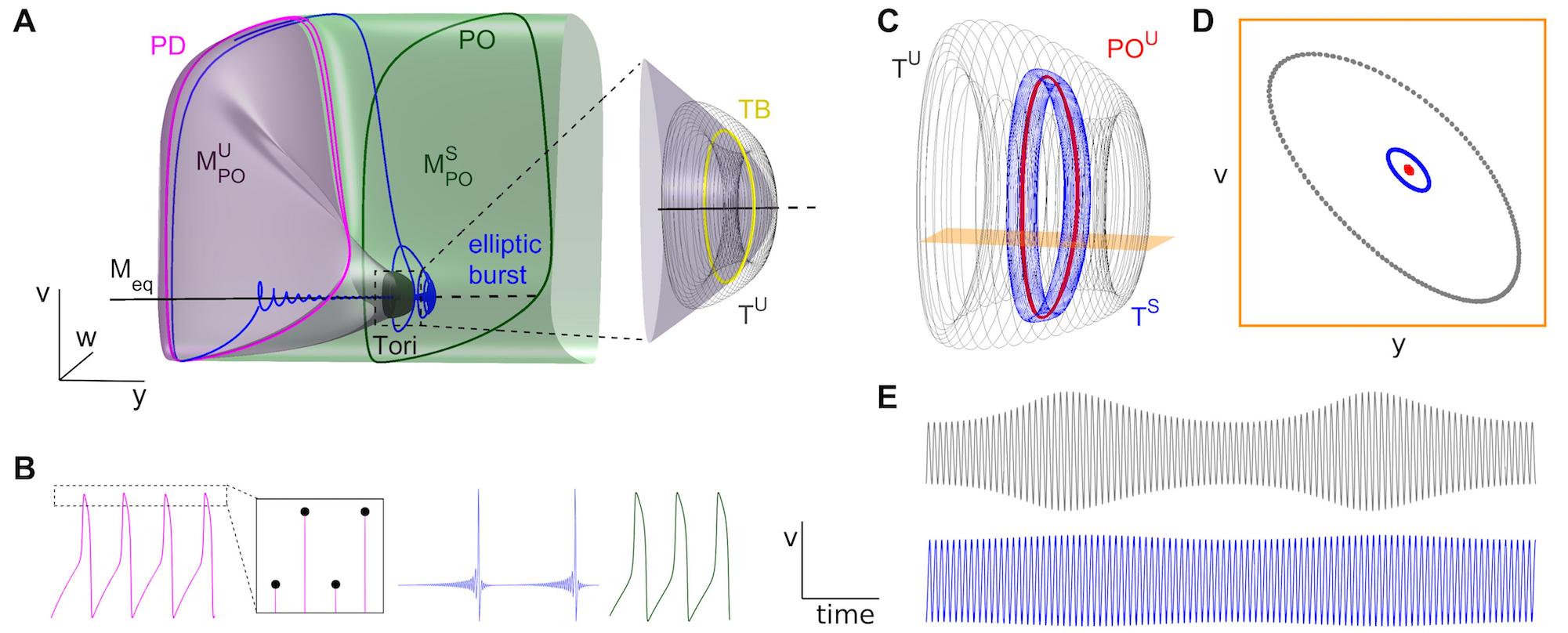}
	\caption{$\mathbf{A}$ 3D ($y,\,w,\,v$)-phase space of the FNR model at $\delta = 0.08$ showing 2D tonic-spiking manifold with its unstable inward (purple surface), M$_{\mathrm{PO}}^{\mathrm{U}}$, merging with stable outward section (green cylinder-shape), M$_{\mathrm{PO}}^{\mathrm{S}}$ at the fold, as well as the 1D quiescent manifold M$_{\mathrm{eq}}$ with stable and unstable (solid/dashed)  branches separated by the sub-critical AH bifurcation. Shown are trajectories of period doubling (magenta circle PD at $c= -0.6191$), elliptic bursting  (blue line at $c =-0.5$) and tonic spiking (dark green circle PO at $c=-0.94$); their corresponding $v$-traces are presented in $\mathbf{B}$. $\mathbf{C}$ 2D repelling torus (grey, $\mathrm{T^{U}}$) containing the nested stable torus (blue, $\mathrm{T^{S}}$) enclosing a repelling periodic orbit (red circle $\mathrm{PO}^{\mathrm{U}}$) at $c=-0.944$. $\mathbf{D}$ 2D cross-section (orange plane) depicting the nested repelling, stable ICs and repelling FP (matching colors in $\mathbf{C}$). $\mathbf{E}$ Slowly modulated voltage traces of the tori in $\mathbf{C}$.  
	} \label{fig:rinzelNemo}
\end{figure*}
It is obvious though that the averaging technique for a 1D slow equation cannot predict the occurrence of the torus bifurcation. Note that the averaging approach illustrated in Fig.~\ref{fig:average}$\mathbf{A_2}$ is done at the parameter values $\mathrm{I_{app}}$ and $\Delta$ when the saddle-torus separates two stable oscillatory states: spiking and bursting. Although the averaging approach is only applicable to detect tonic-spiking state orbits, it can be yet helpful to locate the torus (Fig.~\ref{fig:average}$\mathbf{B_1}$). Actually, the slow averaging applied to the torus orbit in the phase space does reveal the corresponding invariant circle in the $\mathrm{\langle M' \rangle-\langle M \rangle}$ projection as shown in Fig.~\ref{fig:average}$\mathbf{B_2}$.

\section{4D Parabolic burster model}

The same kind of bistability due to a saddle torus at the fold, separating the stable tonic spiking and bursting, can also be observed in a simpler 4D model of parabolic burster. This Hodgkin-Huxley type model is given by
\begin{align}
\nonumber C_{\mathrm{m}}V' = & \ -I_{\mathrm{L}}-I_{\mathrm{K}}-I_{\mathrm{Ca}}-I_{\mathrm{KCa}}-I_{\mathrm{CaS}}+I_{\mathrm{app}},\\
\nonumber n' = & \ \phi(n_{\infty}(V)-n)/\tau_{n}(V),\\
\nonumber \mathrm{[Ca^{2+}]}' = & \ \varepsilon({\mu}I_{\mathrm{Ca}}-[\mathrm{Ca^{2+}}]),\\ 
s' = & \ \varepsilon(s_{\infty}(V)-s)/\tau_{s},  \qquad \varepsilon = 0.0005,  
\label{eq:plant}
\end{align}
(see its full equations in the Appendix below). This model for an endogenously parabolic burster, i.e, with the inter-spike frequency maximized in the middle of each burst,  was proposed and examined in \cite{ermentrout2010mathematical}. Its bifurcation feature is the homoclinic saddle-node bifurcation (SNIC) in the fast 2D subsystem that should demarcate the entry and exit points of the corresponding tonic-spiking and quiescent phases of bursting activity in the full system~(\ref{eq:plant}). Figure~\ref{fig:plant} shows the topology of slow-notion manifolds in the phase-space projection of the model. Here the 2D cylinder-shaped manifold $\mathrm{M_{PO}}$ originates from the depolarized branch of the 1D quiescent manifold  $\mathrm{M_{eq}}$ through a subcritical Andronov-Hopf bifurcation. Next, the unstable (purple) section $\mathrm{M_{PO}^U}$ merges with the stable (green)  section $\mathrm{M_{PO}^S}$ through the saddle-node bifurcation at the first fold. The second fold on  $\mathrm{M_{PO}}$ takes place at the higher values of the $\mathrm{[Ca^{2+}]}$-variable, where the torus bifurcation takes place in the full system. Here, the saddle torus (red) bounds the attraction basin of the tonic-spiking orbit (hidden inside the saddle torus) from that of the stable bursting orbit (black). Inset~\ref{fig:plant}{\bf A} depicts a local transversal cross-section with a repelling IC, representing the saddle-torus, that surrounds a light blue fixed point corresponding to the stable tonic-spiking orbit at the right fold on the manifold $\mathrm{M_{PO}^U}$.

%
The FitzHugh-Nagumo-Rinzel (FNR) system is a formal mathematical neuron model of an elliptic buster \cite{Wojcik2011}:
\begin{align}
\nonumber v' = & \ v-v^3/3-w-y+I_{\rm ext},\\
\nonumber w' = & \ \delta(0.7+v-0.8w),\\
y' = & \ \mu(c-y-v),
\label{eq:FHN}
\end{align}
where $v$ and $w$ represent the fast ``voltage'' and ``gating'' variables, while $y$ is the slow variable due to the smallness of $\mu=0.002$. Here we fixed $I_{\rm ext}=0.3125$ and use the parameter $c$ in the slow equation to continue and reveal the topology of the slow-motion manifolds of this system; the other bifurcation parameter $\delta$ is used to demonstrate how time scales of the variables of the model can determine the bifurcations (torus and period-doubling) of periodic orbits of this model. Figure~\ref{fig:rinzelNemo}\textbf{A} shows the tonic spiking manifold of the FNR model consisting of an unstable inner layer M$_{\mathrm{PO}}^{\mathrm{U}}$ (purple surface) emerging though a subcritical Andronov-Hopf bifurcation from the 1D quiescent manifold, M$_{\mathrm{eq}}$, and a stable outer layer M$_{\mathrm{PO}}^{\mathrm{S}}$ (green surface), both merging though a fold. 

\begin{figure*}[t!]
	\includegraphics[width=.85\textwidth]{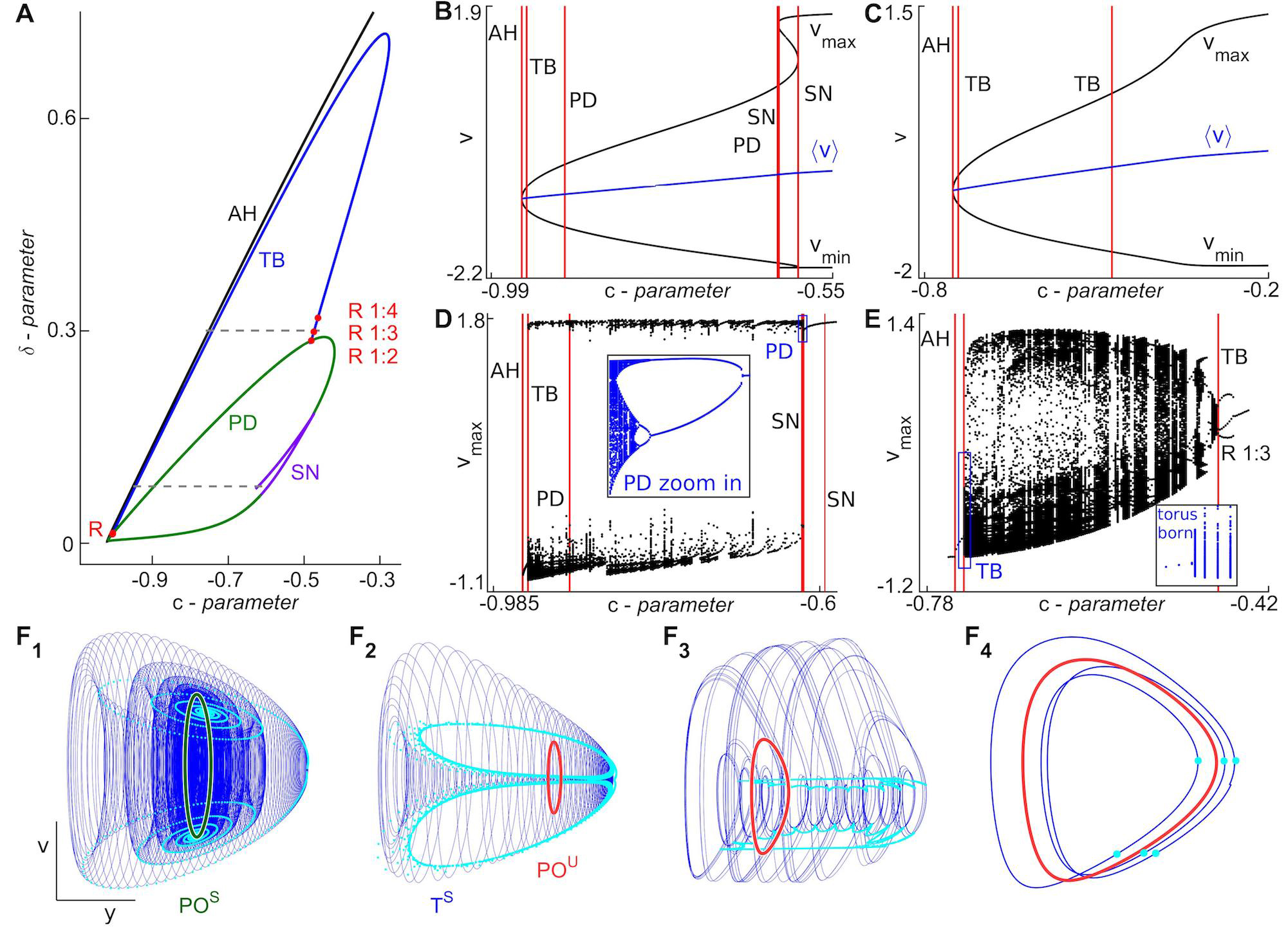}
	\caption{$\mathbf{A}$ Bifurcation diagram of the FNR model depicts the curves for the following bifurcations:  Andronov-Hopf (AH, black line), period-doubling (PD, green line), saddle-node  (SN, purple line) and torus bifurcation (TB, blue line) with indicated strong resonances 1:4, 1:3 and 1:2 (red dots).  Dashed grey lines: c-parameter pathways at $\delta = 0.08$ and $\delta = 0.3$, presented in $\mathbf{B}$--$\mathbf{C}$ and in $\mathbf{D}$--$\mathbf{F}$. $\mathbf{B}$--$\mathbf{C}$ Parameter continuations of $\mathrm{v_{max}}$, $\mathrm{v_{min}}$ and $\mathrm{\langle v \rangle}$ plotted against the c-parameter at $\delta = 0.08$ and $\delta = 0.3$, resp., with red vertical lines indicating ongoing bifurcations. $\mathbf{D}$--$\mathbf{E}$ The c-parameter sweeping diagrams for $\mathrm{v_{max}}$ at $\delta = 0.08$ and $\delta = 0.3$, resp., reveal: $\mathbf{D}$ а PD-cascade en route to elliptic busting, and $\mathbf{E}$ the torus emergence, evolution and breakdown through  1:3-resonance. $\mathbf{F_{1}-F_{4}}$ 2D $\mathrm{(y,\, v)}$-phase projections depicting  a stable PO (green) with a whirlpool born after AH at $c=-0.743$; loss of its stability (red circle) to a stable ergodic torus  (TB at $c =-0.7405$);  a weakly resonant torus at $c = -0.6$ and a 1:3 stable PO at  $c = -0.47$; light-blue highlighted are intersection points of the trajectories with a local transverse plane.}  
	\label{fig:rinzel6a}
\end{figure*}

The $(c,\,\delta)$-bifurcation diagram, shown in Fig.~\ref{fig:rinzel6a}\textbf{A}, of the FNR model is obtained using the parameter continuation package MATCONT \cite{dhooge2003matcont}. It includes several bifurcation curves: AH stands for the Andronov-Hopf [supercritical] bifurcation of the only equilibrium state of the FNR model: TB stands for the torus bifurcation; along this curve the multipliers, $e^{\pm i \phi}$, of the bifurcating periodic orbit runs through strong resonances detected at points labeled as R~1:4 with $\phi=\pi/4$, and R~1:3 with $\phi=2\pi/3$, through the final resonance R~1:2 where $\phi$ completes the arch of the unit circle and reaches $\pi$. The last point is included in the PD-curve corresponding to the period doubling bifurcation through which the periodic orbit looses its stability. The diagram also includes two bifurcation curves, SN, of saddle-node periodic orbits. These curves follow  the PD-curve closely. They cannot be further continued in the $c$-parameter any further down the bifurcation diagram for the given slope of the slow nullcline $v=c-y$ (being a 2D plane in the 3D phase space) because it becomes tangent to the folds of the 2D spiking manifold  M$_{\mathrm{PO}}$. We hypothesize that the bifurcation curves, SN's, and PD, will terminate at the corresponding resonances, 1:1 and 1:2 occurring near the point, labelled as a red dot R (with $\delta \simeq 0$) in the $(c,\delta)$-parameter plane.  Geometrically, this point corresponds to a cusp underlying the transition from the hysteresis to the monotone, increasing dependence of the V${\rm max/min}$- and $\mathrm{\langle V \rangle}$-coordinates of the periodic orbits on the $c$-parameter in the bifurcation diagrams in Figs.~\ref{fig:rinzel6a}({\bf B}) and ({\bf C}), respectively.              

Observe the $\cap$-shape of the TB-curve that indicates that depending on the level of the $\delta$-parameter, the periodic orbit can undergo  two supercritical torus bifurcations, forward and backward, when its looses stability to the torus and re-gains for $\delta=0.3$ [along the upper dashed grey line], or through the forward torus bifurcation followed by the period-doubling bifurcation as $c$ is increased, for $\delta = 0.08$, (lower dashed grey line). 

 The bifurcation diagrams in Figs.~\ref{fig:rinzel6a}\textbf{B}-\textbf{C}, representing the maximum,  $\mathrm{v_{max}}$, and minimum values, $\mathrm{v_{min}}$, of the voltage plotted against the $c$-parameter, illustrate how the tonic-spiking manifold is shaped and where the periodic orbits that constitute it, bifurcate (indicated by vertical red lines) as $c$-parameter is varied for fixed values of  $\delta = 0.08$ and $\delta = 0.3$, respectively. One can clearly see that at a rather large value $\delta = 0.3$ the torus bifurcations do not occur close to the fold of M$_{\mathrm{PO}}$ unlike the period-doubling bifurcation that occurs between two saddle-node bifurcations (indicated by the vertical lines labeled as SNs in Fig.~\ref{fig:rinzel6a}\textbf{B}) corresponding to the two folds on M$_{\mathrm{PO}}$ at $\delta = 0.08$. 

To verify the bifurcations detected by the parameter continuation in the FNR-model, we ran straight-forward $c$-parameter sweeping simulations of the limiting solutions of the model when  $\delta = 0.08$ and when $\delta = 0.3$. The corresponding sweeping diagrams are represented in Figs.~ \ref{fig:rinzel6a}\textbf{D} and \textbf{E}. 

At the level $\delta = 0.3$, after crossing AH-line in the bifurcation diagram Fig.~\ref{fig:rinzel6a}\textbf{A} , trajectories of the FNR-model are attracted into a whirlpool stirred by a stable PO in the phase space shown in Fig.~\ref{fig:rinzel6a}$\mathbf{F_{1}}$. After crossing the left branch of TB-curve, this stable PO loses stability (red circle) to a stable [ergodic and round] torus in the phase space depicted in Fig.~\ref{fig:rinzel6a}$\mathbf{F_{2}}$. The sweeping bifurcation diagram in Fig.~\ref{fig:rinzel6a}\textbf{E} shows that with a further increase of $\mathrm{c}$, the torus becomes  resonant (Fig.~\ref{fig:rinzel6a}$\mathbf{F_{3}}$) and its shape becomes more complex giving rise to the so-called mixed-model oscillations (MMO) of alternating small-amplitude sub-threshold and large spiking ones.  Once crossing the right branch of TB line, the torus goes into a strong resonance 1:3 before it breaks down at larger values of the $c$-parameter giving rise to the stable period-3 orbit in the phase space, see  Fig.~\ref{fig:rinzel6a}$\mathbf{F_{4}}$.

At $\delta = 0.08$, after crossing the left branch of TB-curve in the bifurcation diagram in  Fig.~\ref{fig:rinzel6a}\textbf{A}, the PO loses stability (red line in Fig.~\ref{fig:rinzelNemo}\textbf{C}) to a stable torus (blue line) in the phase space. Interestingly, at $\mathrm{c = -0.94415}$, this small stable torus bounding the unstable PO is nested inside a larger repelling  unstable torus (grey line). The
coexistence of the both tori is better seen in a local Poincar\'{e} cross-section (an orange plane) chosen transversally to these orbits  as shown in Fig.~\ref{fig:rinzelNemo}\textbf{D}). Note that in a 3D system a repelling torus can be detected and traced down in the backward time when it becomes attracting, which is impossible for saddle tori occurring in high-dimensional systems.   

\begin{figure}[t!]
	\includegraphics[width=.9\columnwidth]{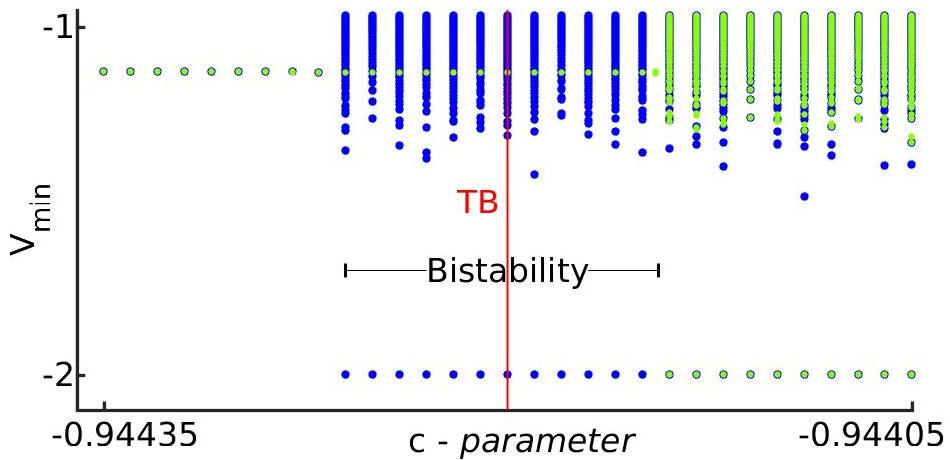}
	\caption{Sweeping diagram of the attractors in the FNR model, represented by their $\mathrm{V_{min}}$-values (green/blue dots) plotted against the $c$-parameter being increased/decreased. Within the bistability window (horizontal black bar), tonic-spiking and bursting attractors co-exist and are separated by the repelling torus. Vertical (red) line at $c = -0.944145$ indicates the torus bifurcation detected in MATCONT.}
	\label{fig:FNRbi}
\end{figure}

 The same approach employed to detect bistability in the Purkinje cell model due to a hysteresis effect, is used  to visualize bistability in the FNR-model as its solutions are swept in the opposite directions of the $c$-parameter. The obtained bifurcation diagram, with $v_{\rm min}$-values (as green/blue dots) plotted against increasing/decreasing $c$-values, is represented in Fig.~\ref{fig:FNRbi}. Here a single  dot for the given $c$-value corresponds to a stable periodic orbit of a single $v_{\rm min}$-value, while several vertical dots correspond to co-existing busting orbits. The basins of these orbits are bounded by the  repelling torus enclosing the stable tonic spiking orbit in the 3D phase space of the FNR-model. 
 
\begin{figure*}[t!]
	\includegraphics[width=.8\textwidth]{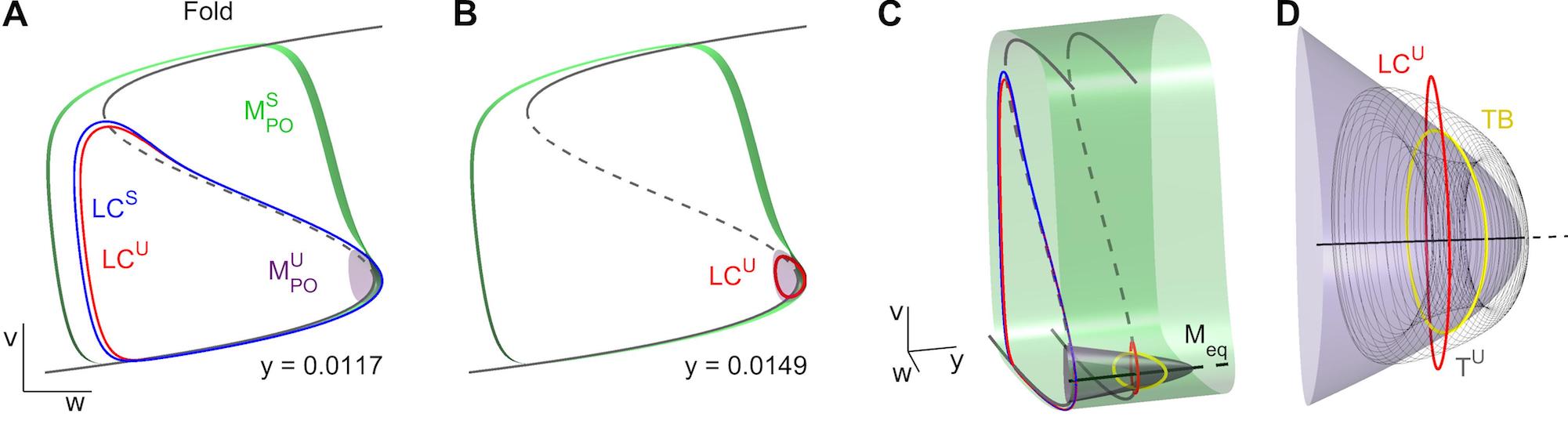}
	\caption{$\mathbf{A}$ 2D $(v,\,w)$-projection depicting  the stable (green) M$_{\mathrm{PO}}^{\mathrm{S}}$, and unstable (purple) M$_{\mathrm{PO}}^{\mathrm{U}}$ branches of the tonic spiking-spiking manifold of the FNR system at $\delta = 0.08$ that are superimposed $\mathbf{A}$ with the saddle-node limit cycle (which then bifurcates into a stable (blue) limit cycle LC$^{\mathrm{S}}$ and an unstable (red) limit cycle LC$^{\mathrm{U}}$) of the fast subsystem at $y = 0.0117$, and $\mathbf{B}$ with limit-cycle canards following the stable and unstable branches of the cubic fast nullclines at $y = 0.0149$. $\mathbf{C}$ 3D phase space depicting M$_{\mathrm{PO}}^{\mathrm{S}}$ and M$_{\mathrm{PO}}^{\mathrm{U}}$ being superimposed with stable and unstable LC-canards from $\mathbf{A}$--$\mathbf{B}$. Inset $\mathbf{D}$ zooms-in the torus ($\mathrm{T^{U}}$, grey), the unstable limit cycle LC$^{\mathrm{U}}$ (red ring) around $y = 0.0149$ and the torus bifurcation (TB) detected by MATCONT (yellow ring). }  
	\label{fig:rinzelSN}
\end{figure*}

 \begin{figure}[t!]
 	\includegraphics[width=.9\columnwidth]{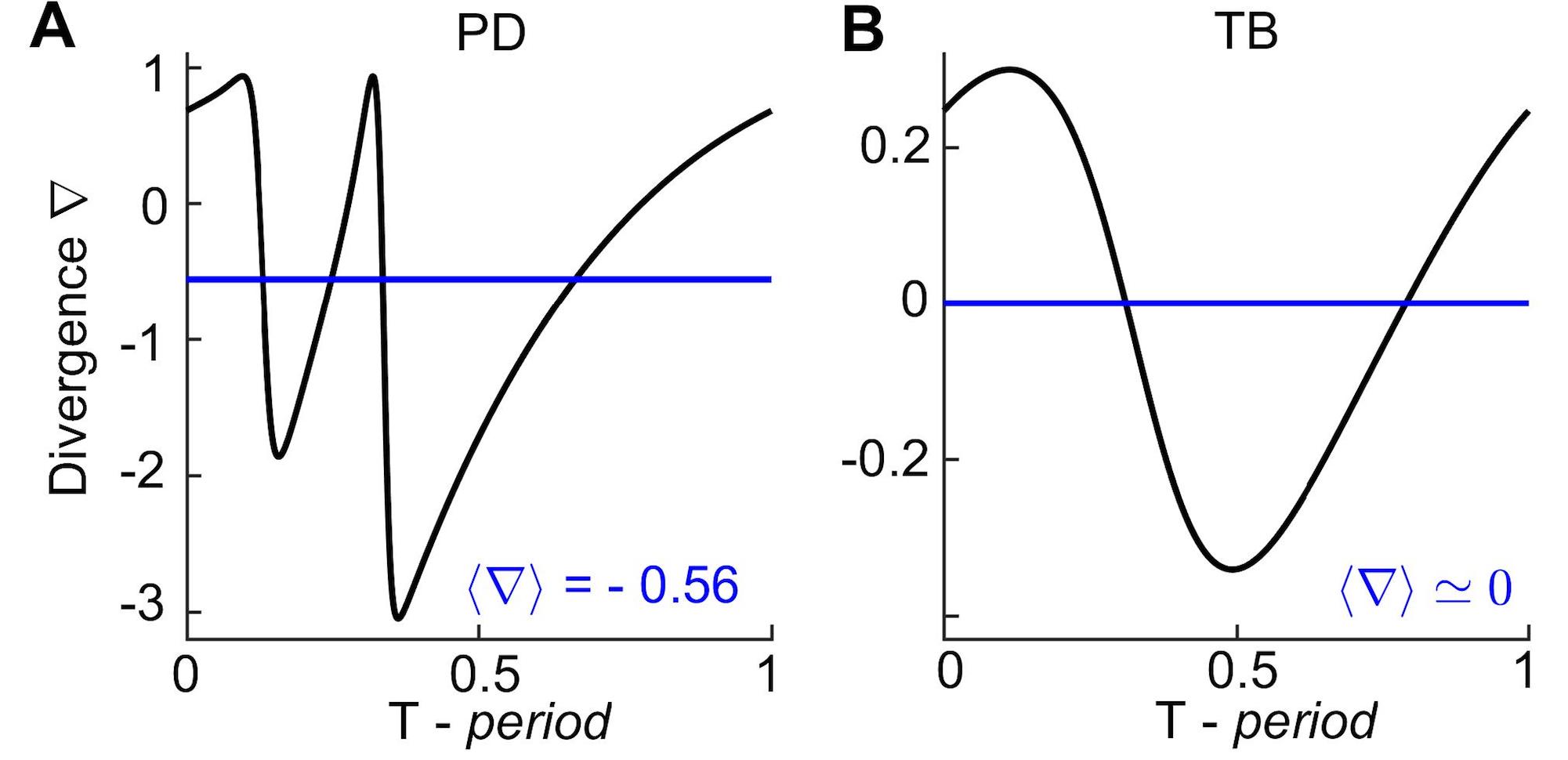}
 	\caption{Divergence $\nabla = \left [ \frac{\partial v'}{\partial v}+\frac{\partial w'}{\partial w}+\frac{\partial y'}{\partial y} \right ]$ along the periodic orbit (with period normalized to 1) superimposed with the blue level-curve for the average divergence $\langle\nabla\rangle$ (its value shown below) for the PD-bifurcation scenario (\textbf{A}) or for the TB scenario (\textbf{B}) at $\delta = 0.08$.}
 	\label{fig:FHNdiver}
 \end{figure}

 Again we note that the tori emerge in the FNR model near  the smooth cone-shaped end of M$_{\mathrm{PO}}^{\mathrm{U}}$ instead of around its fold (Fig.~\ref{fig:rinzelNemo}\textbf{A} inset). Fig.~\ref{fig:rinzelNemo}\textbf{E} shows the slowly-modulated ``voltage'' traces corresponding to the two tori. As $c$-parameter further increases to $\mathrm{c = -0.94}$, elliptic bursting becomes dominantly stable in the phase space shown in Figs.~\ref{fig:rinzelNemo}\textbf{A} and \textbf{B}). The period-doubling cascade starts at the left branch of the PD-curve and ends at its right branch in the bifurcation diagram in Fig.~\ref{fig:rinzel6a}\textbf{A} and Fig.~\ref{fig:rinzel6a}\textbf{D} (magenta line in Fig.~\ref{fig:rinzelNemo}\textbf{A} and \textbf{B}, $\mathrm{c = -0.6191}$). After that, tonic spiking (PO in phase space) becomes stable (dark green line in Fig.~\ref{fig:rinzelNemo}\textbf{A} and \textbf{B}, $\mathrm{c = -0.5}$). Note that in the FNR model, PD instead of TB occurs at the fold.

\subsection{Origin of slow-fast torus bifurcation}
Let us outline a possible origin and stability of the torus bifurcation in the FNR-model. Its start with a saddle-node [fold] bifurcation of two limit cycles, stable and repelling, that occurs in and due to the 2D fast $(v,\,w)$ sub-system of the model, i.e., whenever the slow $y$-variable is frozen and used as a parameter. So, the value $y = 0.00117$ is the critical one at which the stable (blue circle) $\mathrm{LC^{S}}$ and unstable (red circle)  $\mathrm{LC^{U}}$ limit cycles emerge through a saddle-node [fold] bifurcation in the $(v,\,w)$-subspace, see Fig.~\ref{fig:rinzelSN}\textbf{A}. While the stable limit cycle $\mathrm{LC^{S}}$  increases in its size to become the large-amplitude orbit of the relaxation oscillator, the unstable limit cycle $\mathrm{LC^{U}}$ keeps shrinking as $y$ increases, see the corresponding $(v,\,w)$-phase portrait in Figs.~\ref{fig:rinzelSN}\textbf{B} and \textbf{C}. As the result, in the full model, the slow increase/decrease of the $y$-variable makes the $\mathrm{LC^{U}}$ drag back and forth to form the 2D repelling torus in the 3D phase space of the model, see Fig.~\ref{fig:rinzelSN}\textbf{D}.   
%
\begin{figure*}[t!]
	\includegraphics[width=.7\textwidth]{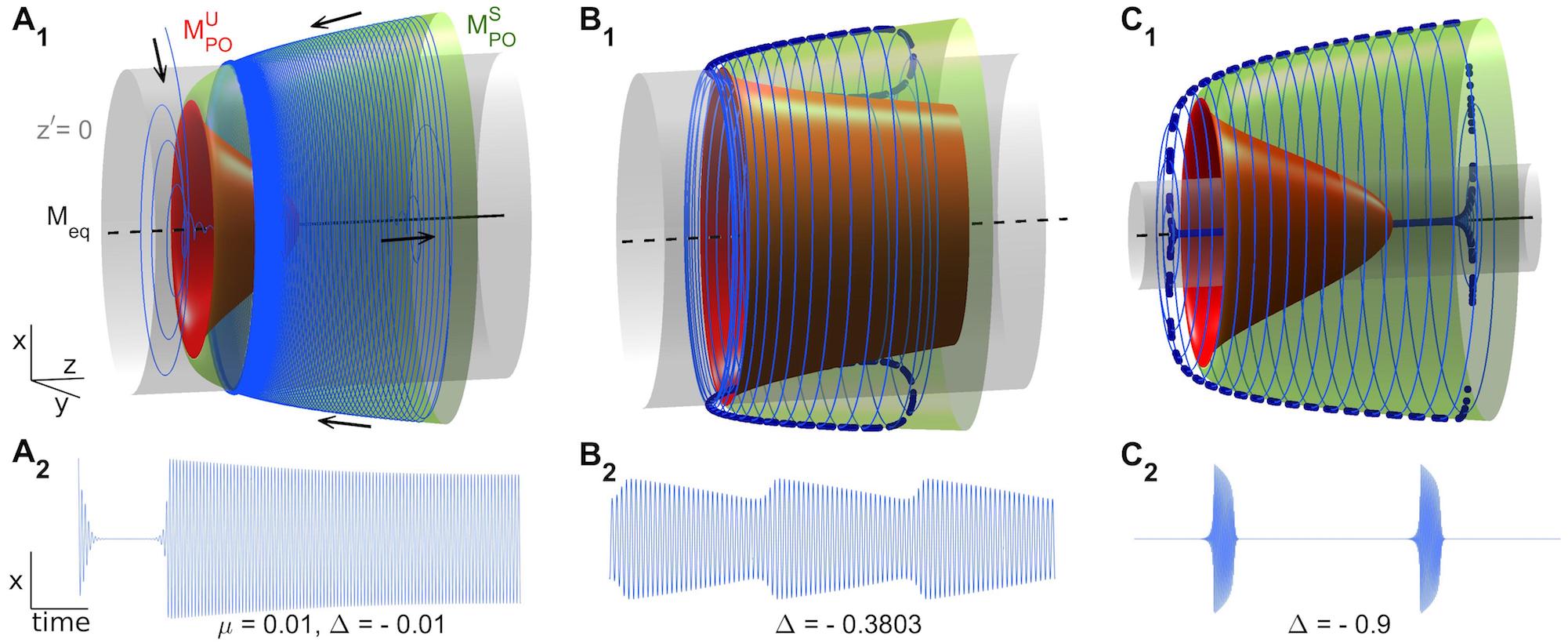}
	\caption{The $(z,\, y,\, x)$-phase space of the the 2-layer model~(\ref{eq:NF}) depicting the tonic-spiking manifold with an outward stable part M$_{\mathrm{PO}}^{\mathrm{S}}$ (green surface) merging with the inward  unstable section M$_{\mathrm{PO}}^{\mathrm{U}}$ (red surface) at the fold, and the quiescent manifold M$_{\mathrm{eq}}$ (black line) with stable (solid) and unstable sections on the $z$-axis. the slow nullcline $z' = 0$ is represented by the grey cylinder as its radius at selected  $\Delta$-values: -0.01, -0.3803, -0.9. An outside trajectory (blue line) in \textbf{A$_{\mathrm1}$} quickly spirals onto M$_{\mathrm{eq}}$, then gets ``sucked'' into the cylinder (where $z'>0$), and after a delayed loss of the stability past the subcritical AH point, spirals away towards M$_{\mathrm{PO}}^{\mathrm{S}}$ outside the cylinder where ($z'<0$), on which it converges to a periodic orbit (black arrows indicates the trajectory direction). The stable torus-canard in \textbf{B$_{\mathrm1}$}  transitions into an elliptic burster in \textbf{C$_{\mathrm1}$}. Other parameters: $\omega=2,\; l_{1}=1,\; l_{2}=-0.8$. Insets \textbf{A$_{\mathrm2}$}--\textbf{C$_{\mathrm2}$} show the corresponding ``voltage'' traces.}  
	\label{fig:NF}
\end{figure*}

\subsection{Divergence-sign test to predict TB or PD}      
Either the torus bifurcation (TB) or the period-doubling bifurcation (PD) can underly the stability loss of the periodic orbit existing on the attracting cylinder-shape section of the tonic-spiking manifold M$_{\mathrm{PO}}$ in the phase space of the FNR-model when it reaches the fold, see Fig.~\ref{fig:rinzelNemo}. We have already noticed depending on the value of the $\delta$-parameter, this large amplitude orbit losses stability though the torus bifurcation followed by its resonances and breakdown for $\delta \ge 0.3$ and or via a period doubling bifurcation at smaller values like $\delta = 0.08$. This is indicative that wherever the time scales of the $v$- and $w$-variables becomes of a similar order the torus bifurcation case prevails over the period doubling one where both $w,\,y$-variables become much slower than the $v$-variable, and vice versa.  
%
\begin{figure*}[t!]
	\includegraphics[width=.7\textwidth]{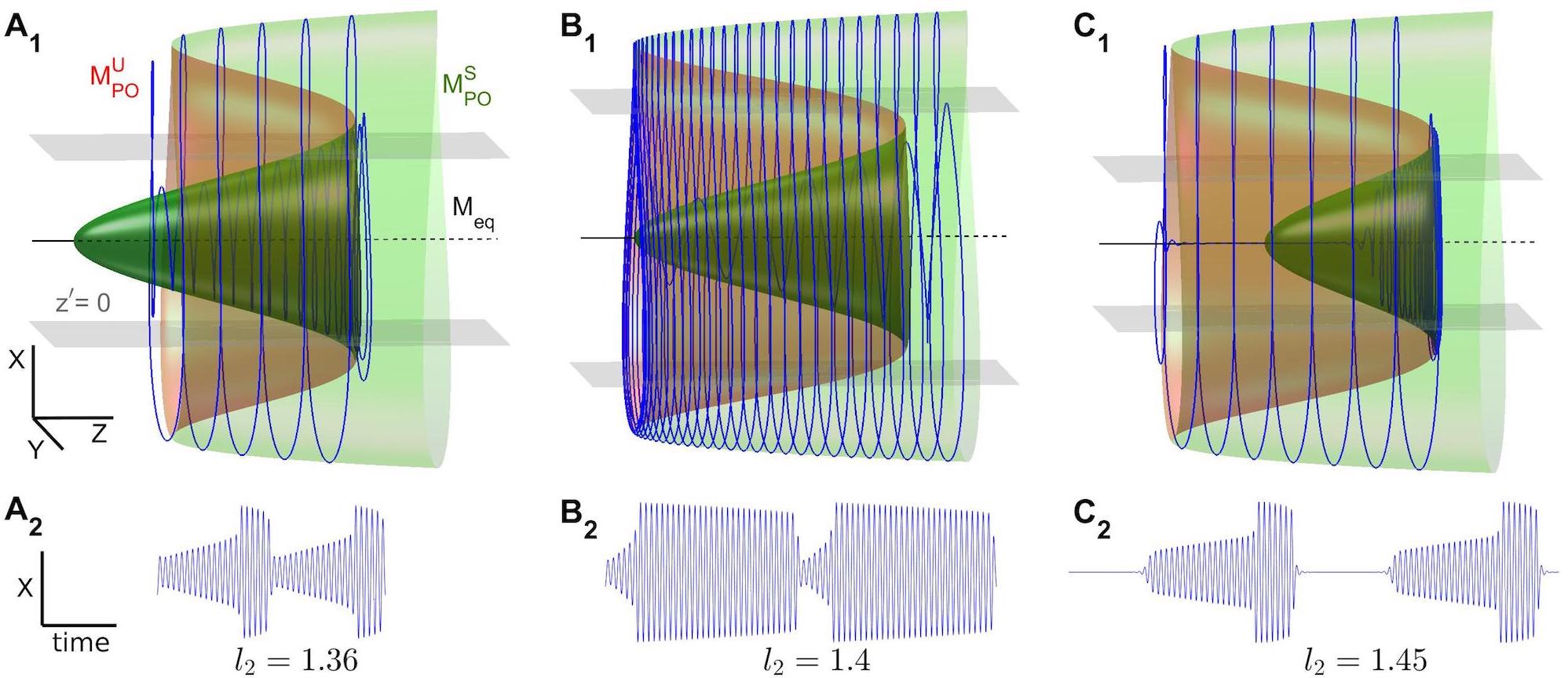}
	\caption{The $(z,\, y,\, x)$-phase space of the 3-layer [relaxation torus] model ~(\ref{eq:NFthree}) showing the tonic-spiking manifolds with two folds connecting its inner and outer stable (green surfaces, M$_{\mathrm{PO}}^{\mathrm{S}}$) and the middle unstable (red surface, M$_{\mathrm{PO}}^{\mathrm{U}}$) branches. The quiescent manifold, the $z$-axis (black line) has two sections: stable ($z<0$, solid) and unstable ($z>0$, dashed). The slow nullcline $z' = 0$ is a pair of parallel (grey) planes between which $z' > 0$, and $z' < 0$ outside.     
		\textbf{A$_{\mathrm1}$}--\textbf{C$_{\mathrm1}$} Rearrangements of M$_{\mathrm{PO}}$ with $l_{2}$-variations at indicated values reshape modulations of voltage traces shown in \textbf{A$_{\mathrm2}$}--\textbf{C$_{\mathrm2}$} at $\Delta$ = 1.5, 3.5 and 1.092; other parameters: $l_{1} = -4.9, l_{3} = -0.1, \omega = 10,\, \mu = 0.5$.}
	\label{fig:nf3model}
\end{figure*}

To predict the type of the bifurcation, torus or period doubling, the stable orbit will undergo, without computing its multipliers, we can evaluate the sign of the divergence given by $\nabla =\left  [ \frac{\partial v'(t)}{\partial v}+\frac{\partial w'(t)}{\partial w}+\frac{\partial y'(t)}{\partial y} \right  ]$ 
 along the periodic orbit on the fold of the slow-motion manifold in the 3D phase space of the FNR  model. The arguments are similar to the well-known Bendixson-Dulac criterion for the existence of a close orbit, like a limit cycle and a separatrix loop of a saddle in a system on a 2D plane; note the Dulac's argument follows from the Green's theorem \cite{king2003differential}. Loosely speaking, it can be directly iterated as follows: no closed orbit exists in a domain of the 2D phase plane where the vector field divergence is of a constant sign. As far as the 2D torus in a 3D phase space is concerned, it can not emerge in a 3D subspace where the divergence does not change its sign. For example, this guarantees no torus bifurcation in  the classical Lorenz-63 model with the chaotic attractor, which is a dissipative system with a constant negative divergence to collapse phase volumes. On the other hand, the torus bifurcation occurs in the 3D Lorenz-84 model \cite{shil1995bifurcation}.  One of the features of the Lorenz-84 model is that the divergence of its vector filed is phase-coordinate dependent and hence can be sign-alternating. This is why the torus bifurcation is a part of bifurcation unfolding of an equilibrium state with the characteristic exponents $(0, \pm i \omega)$ (the sum of the exponents vanishes), which is also referred to as the codimension-two Gavrilov-Guckenheimer or simpler fold-Hopf bifurcation, which is the key feature of the Lorenz-84 model. Such a bifurcation may also occur in the FNR-model near the tip of the cone-shaped manifold M$_{\rm PO}$ crossed out by the slow nullcline $y'=0$ with $\mu \ll 1$.

Thus, if the averaged divergence, $\langle\nabla\rangle$, i.e., all $\nabla$-values summed up along the periodic orbit and  averaged over its period, on the fold of the manifold M$_{\rm PO}$, is close to zero then the stable periodic orbit may undergo the torus bifurcation at the stability loss. Otherwise if $\langle\nabla\rangle <0$, which is typical for most 3D dissipative systems, the periodic orbit will loose stability through the period-doubling bifurcation. The calculation of $\langle\nabla\rangle$ in the FNR model~(\ref{eq:FHN}) at various points on the PD- and TB-curves supports our hypothesis. Figure~\ref{fig:FHNdiver} shows how $\nabla$ varies over the normalized period of the orbit loosing the stability though the period-doubling and torus bifurcations. Its average  $\langle\nabla\rangle$-value is about $\sim 0.56$ and  0 in the period-doubling and torus cases, respectively, when $\delta = 0.08$. This {\em de-facto} validates the divergence-based approach and that it can be used to predict the type of the bifurcations will occur at the loss of stability of periodic orbits on the distinct fold of the slow-motion manifold in the phase space of 3D slow-fast systems.   

 \begin{figure*}[ht!]
 	\includegraphics[width=.85\textwidth]{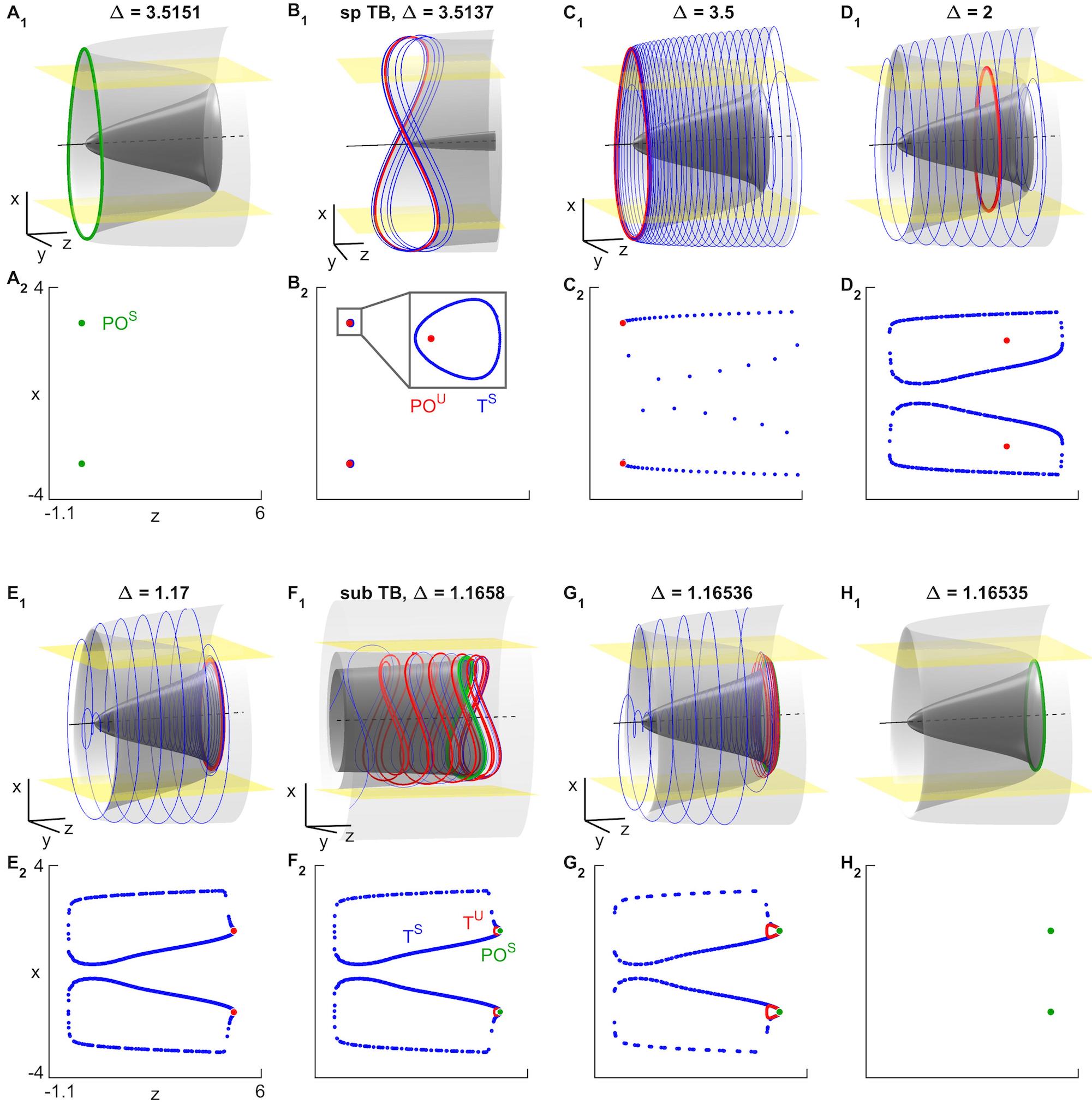}
 	\caption{Torus-canard transformations in the 3-layer [relaxation-torus] model~(\ref{eq:NFthree}) as $\Delta$-parameter is decreased at indicated values. 3D $(z, y, x)$-phase space depicts the tonic-spiking manifold M$_{\rm PO}$ (grey surface) with two folds connecting its stable inner, outer and middle unstable sections.  The quiescent manifold is the $z$-axis (black line) with stable ($z<0$, solid) and unstable ($z>0$, dashed) sections. The slow nullcline $z' = 0$ is a pair of parallel (yellow) planes between which $z' > 0$.   \textbf{A$_{\mathrm1,2}$} -- \textbf{E$_{\mathrm1,2}$} Stable PO (green circle) loses stability to a torus-canard (blue IC on a Poincar\'e cross-section given by $y=0$) near the left fold that becomes wider and weekly resonant.   \textbf{ F$_{\mathrm1,2}$}--\textbf{H$_{\mathrm1,2}$} Unstable PO (red cycle/FP) regains stability with a sub-critical TB generating a small repelling torus (red IC, $\mathrm{T^{U}}$) that grows in size, merges with the stable torus-canard $\mathrm{T^{S}}$ and both annihilate, leaving the stable PO on the inner branch of  M$_{\rm PO}$. Other parameters: $l_{1} = -4.9, l_{2} = 1.4, l_{3} = -0.1, \omega = 10, \mu = 0.5.$}
 	\label{fig:NFthree}
 \end{figure*}
 %

\section{Normal forms for torus-canards}
The amplitude modulation phenomenon that happens in various complex and realistic neuron models inspires us to build a simple and adjustable mathematical model to generate and analyze torus bifurcations using evident geometric constrains. Since such tori emerge near the fold of the tonic-spiking manifold, the first step is to create the desired number of such folds in the 2D normal-form fast subsystem given by   
\begin{align}
\nonumber x' = & \ ~~~\omega y+x Q(x,y,\varepsilon),\\
          y' = & \ -\omega x+y Q(x,y,\varepsilon)  
          \label{eq:NFgeneral}
\end{align}
with a polynomial function $Q=\varepsilon+\sum_{n=1}^{\infty} l_n(x^2+y^2)^n$. When $Q \equiv 0$, Eqs.~(\ref{eq:NFgeneral})  describe a linear harmonic oscillator that has a single equilibrium state -- the center at the origin in the $(x,\,y)$-phase plane. 
When $Q=\varepsilon<0$, this linear system has a stable focus at the origin, which becomes unstable when $\varepsilon>0$. When $Q(x,y,\varepsilon)$ is negatively/positively defined, the nonlinear 
system has yet a single equilibrium state that globally attracts/repels all other spiral-shaped trajectories. A level curve given by $Q=0$ correspond to a closed orbit -- limit cycle of the system. In case of a single zero (other then the origin), the limit cycle is double or of the saddle-node type. In case of two distinct zeros, they correspond to two nested limit cycles surrounding the origin that determines their stability. So, if the origin is stable for $\varepsilon <0$, then the inner LC is repelling, whereas the outer one is stable, or vice versa if $\varepsilon >0$. In the special case  $\varepsilon =0$, the stability of the limit cycle is determined by the sign of the corresponding Lyapunov coefficient $l_{n}$.


This basically describes the algorithm for devising the desired systems. Let us first consider the model with a single fold. Its equations are given below:      
\begin{align}
\nonumber x' = & \ ~~~\omega y+x\left(z+l_1(x^2+y^2)+l_2{(x^2+y^2)}^2\right),\\
\nonumber y' = & \ -\omega x+y\left(z+l_1(x^2+y^2)+l_2{(x^2+y^2)}^2\right),\\
z' = & \ \mu\left(1-x^2-y^2+\Delta\right), \label{eq:NF}
\end{align}
with $(x,y)$ being the fast variable and $z$ being a slow (as $\mu$ is small), dynamic variable replacing fixed $\varepsilon$. When $\mu=0$, the first two equations present a normal form for the Bautin bifurcation of the equilibrium state at the origin in case when $\varepsilon=0$ and first Lyapunov coefficient, $l_1=0$, of the cubic terms (note that $l_2$ can be scaled to equal $-1$, for simplicity). 
The last constrains implies that the outer LC will be stable and a nested inner one will be repelling as $l_1>0$. The equilibrium state at the origin is stable as long as $z<0$, and becomes unstable when $z \ge 0$ after the repelling LC collapses into it through a sub-critical AH bifurcation at $z=0$. This ends up bi-stability in the fast subsystem system and the stable LC is the only global attractor. The left boundary for bistability is given by $z_{\rm sn}=-\left [ l_1(x^2+y^2)+l_2{(x^2+y^2)}^2 \right ]$ that correspond to the double, saddle-node LC in the $(x,\,y)$-plane. This double orbit vanishes for 
smaller negative $z$-values, making the origin the only global attractor in the phase space.       

The slow equation in (\ref{eq:NF}) is designed here as follows: the slow nullcline, where $z'=0$ is a cylinder given by $x^2+y^2=1+\Delta$ of radius $\sqrt{a+\Delta}$ in the 3D phase space of the system. Inside this cylinder $z'>0$, and  $z'<0$ outside of it. The use of $\Delta$ is two-fold here: as a bifurcation and sweeping parameter to locate the 2D manifold M$_{\rm PO}$ in MATCONT.   

Now, let us put all components together in the full system. Figure~\ref{fig:NF} shows the 3D phase space of the 2-layer model~(\ref{eq:NF}). Here is the 1D quiescent manifold $\mathrm{M_{eq}}$ (the $z$-axis) when stable ($z<0$) and unstable $(z>0)$ branches. The subcritical AH bifurcation gives rise to the unstable (red) branch $\mathrm{M_{PO}^{U}}$ that folds back and continues further (rightwards) as stable (green) branch $\mathrm{M_{PO}^{S}}$. The slow nullcline is represented as a grey cylinder parallel to the $z$-axis. Let us pick an initial point to the left quite far away from the fold on $\mathrm{M_{PO}^{U}}$ in the 3D phase space. Then, the trajectory fast spirals onto the stable branch (solid line) of M$_{\rm eq}$ inside the slow nullcline/cylinder along which it will be slowly dragged to the right toward the AH bifurcation in the fast subsystem. Passing throughout it, the phase point keeps following the unstable branch of $\mathrm{M_{eq}}$ (the so-called delayed loss of stability) before it spirals away to converge to the stable, outer branch $\mathrm{M_{PO}^{S}}$. Provided that radius of $\mathrm{M_{PO}^{S}}$ is larger than that of the slow nullcline, the phase point, turning around $\mathrm{M_{PO}^{S}}$, will slide leftward toward the fold on the 2D spiking manifold. If $\Delta$ controlling the radius of the slow nullcline is large enough, then the phase point will converge to a stable and round periodic orbit on $\mathrm{M_{PO}^{S}}$. Varying $\Delta$ makes this period orbit slide along M$_{\rm PO}$ thus revealing its shape, including the fold. This is the essence of the parameter continuation approach to localize slow-motion manifolds of the full system by varying the control [faux] parameter of the slow subsystem in the phase space. By decreasing $\Delta$, the stable PO slides down towards the fold below which the orbit continues as repelling both unstable in $z$-direction and in the fast ($x,\,y$)-coordinates. As we pointed earlier, the transverse intersection of M$_{\rm PO}$ with the slow nullcline/cylinder guarantees that the PO does not vanish but loses stability at the fold through a super-critical torus bifurcation that result in the emergence of the stable torus-canard (provided $\mu$ is small enough)   switching back and forth between the stable and unstable branches of the tonic spiking manifold near its fold, see Figure~\ref{fig:NF}${\mathbf  B_1}$. Further decreasing $\Delta$ makes the stable torus morph into the ``elliptic'' buster with amplitude modulations and episodes of quiescent meta-states.  \\     

Our last 3-layer (torus-relaxation) model below
\begin{align}
\nonumber x' = & \ ~~~\omega y+x\Big(z+l_1(x^2+y^2)\\
\nonumber &+l_2{(x^2+y^2)}^2+l_3{(x^2+y^2)}^3\Big),\\
\nonumber y' = & -\omega x+y\Big(z+l_1(x^2+y^2)\\
\nonumber &+l_2{(x^2+y^2)}^2+l_3{(x^2+y^2)}^3\Big),\\ 
z' = & \ \mu(1-x^2+\Delta) \label{eq:NFthree}
\end{align}
is designed to have two folds between two stable, inner ($l_1<0$) and outer ($l_3<0$)  and middle unstable ($l_2>0$) branches of the 2D tonic spiking manifold M$_{\rm PO}$, see its phase space in Fig.~\ref{fig:nf3model}. The distinction of the slow equation in Eqs.~(\ref{eq:NFthree}) is that its slow nullcline is represented by two parallel planes $x=\pm \sqrt{1+\Delta}$. The planes, located below and above the $z$-axis, functionally provide the same slow dynamics and means of its control by varying $\Delta$ in the full system.  

Like in the 2-layer model, by varying $\Delta$ one can set the 3-layer model to demonstrate various tonic spiking oscillations and torus-canards. In addition, by calibrating values of  $l_1$ and $l_2$ one can shifts the relative positions of the folds to shape and produce various types of elliptic bursters generated by this system, as Fig.~\ref{fig:nf3model} demonstrates.  Let us describe the torus-canard transformations as $\Delta$ is decreased. The principal stages of the transformation are documented in Fig.~\ref{fig:NFthree}.  

At the initial stage the stable PO exists on the outer surface $\mathrm{M_{PO}^S}$ at some large enough $\Delta$, see Fig.\ref{fig:NFthree}\textbf{A}. 
With an initial decrease, this stable PO slides down off the left fold and loses stability to a torus-canard, $\mathrm{T^S}$ (Fig.\ref{fig:NFthree}\textbf{B}). With further decreasing $\Delta$ the repelling PO slides down along the middle branch $\mathrm{M_{PO}^U}$ to the right fold, meanwhile the stable relaxation torus-canard with a head grows in size expanding the whole space between the inner and the outer layer, its stages are shown in Figs.~\ref{fig:NFthree}\textbf{C}-\textbf{E}. 
 Next, via a subcritical TB, a repelling torus, $\mathrm{T^U}$, emerges at the right fold from the PO that regains the stability, see Figs.~\ref{fig:NFthree}\textbf{F}-\textbf{G}. At this point there are two tori: stable (blue) and unstable (red IC in the cross-section).
     Note that the repelling torus separates the attraction basins of the stable torus from the stable PO, thereby creating bistability in the system. Finally, the stable and the unstable tori merge and annihilate through a saddle-node bifurcation so that  the stable PO at the right fold remains the only attractor in the phase space, see Fig.~\ref{fig:NFthree}\textbf{H}.

\section{Conclusion}

Following the bottom-up approach we have considered and analyzed a collection of chosen exemplary slow-fast models, starting off with the biologically plausible Hodgkin-Huxley ones up to light mathematical toy systems that all feature the torus bifurcations occurring on the characteristic fold on the slow-motion tonic-spiking manifold. Unlike the flat canards occurring in 2D slow-fast systems whose stability can be evaluated analytically as it is dictated by the analytical properties of the function on the right-hand side of the fast equation in the singular limit, the question concerning stability of emergent tori is yet to be fully understood. We have shown all kinds of tori in the model list, from stable and repelling (made stable in the backward time in 3D systems) to saddle ones with 3D unstable manifolds and XD stable manifold, where X is determined by the phase space dimension of the model in question. 
The fact that the torus emerges locally next to the fold of a low-dimensional surface let one use one's skills to choose a suitable Poincar\'e cross-section and to find two initial conditions on it to demonstrate that one trajectory goes inside such a torus to converge to a nested stable periodic, while the other converges to another attractor in the phase space. This bistability is a de-facto proof of the existence of 2D saddle tori in the phase space, and can be further supported by bi-directional parameter sweeps to reveal the hysteresis due to overlapping interval of the co-existence of two attractors, like tonic-spiking and bursting orbits in the Purkinje cell model and the FNR-model.  

\begin{figure}[t!]
	\includegraphics[width=.8\columnwidth]{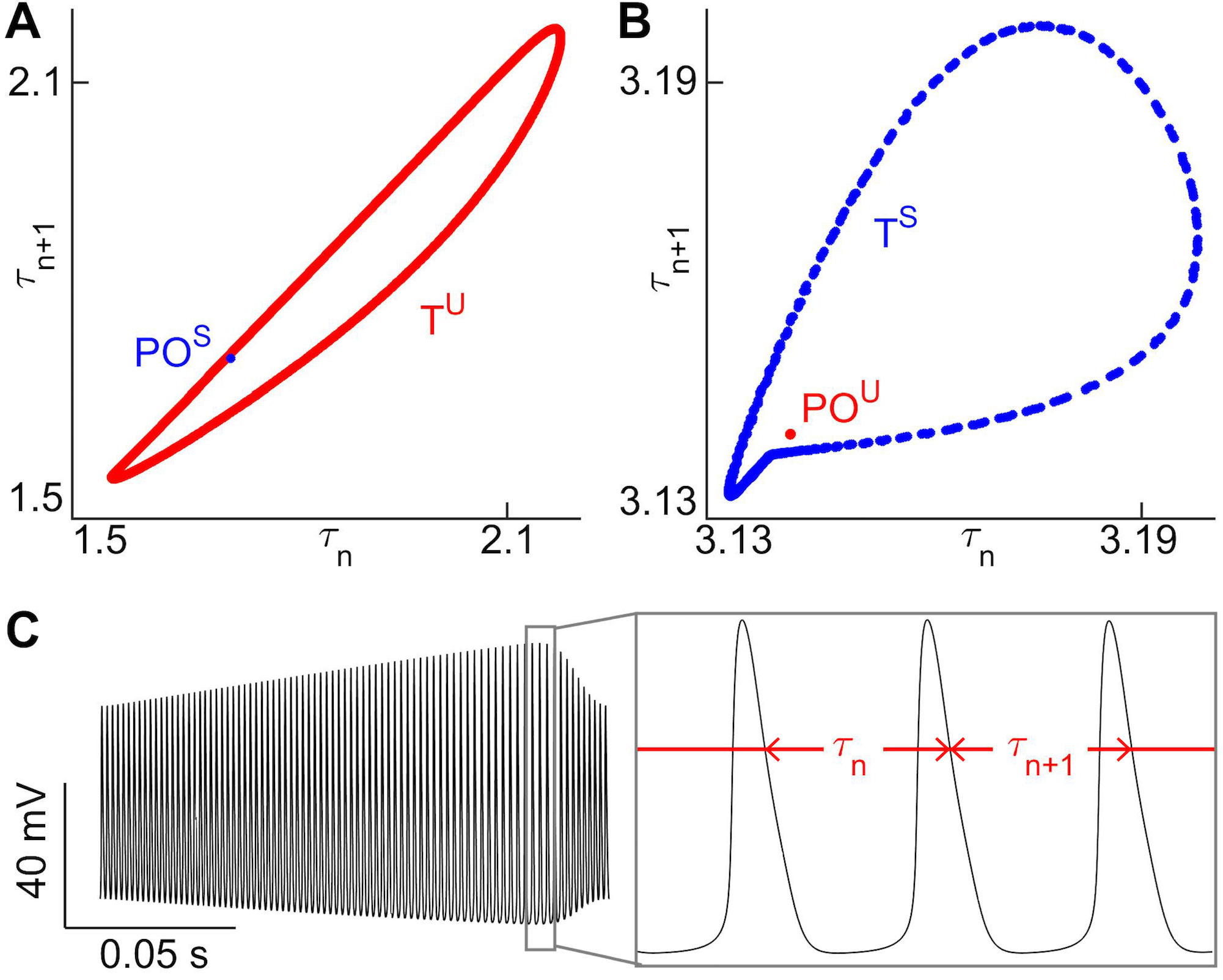}
	\caption{Poincar\'e return maps, $\tau_{n} \to  \tau_{n+1}$, for recurrent time intervals between consecutive spikes in the voltage traces. $\mathbf{A}$ Unstable (red) IC [corresponding to the saddle torus-canard $\mathrm{T^{U}}$] encloses a stable (blue) FP (corresponding to tonic-spiking orbit $\mathrm{PO^{S}}$) in Purkinje cell model at $I_{app}=-29.487$. $\mathbf{B}$ Stable (blue) IC ($\mathrm{T^{S}}$) bounding the unstable (red) FP ($\mathrm{PO^{U}}$) in the  normal-form~(\ref{eq:NF}). Inset $\mathbf{C}$ shows a quasi-periodic voltage trace with recurrent time intervals $\tau_{n}$ employed in the return map in {\bf A} for Purkinje cell model.}  
	\label{fig:ftTwo}
\end{figure}

While parameter continuation packages like  MATCONT and AUTO are very handy to detect torus bifurcations and some strong resonances on the corresponding bifurcation curves, one has to use additional computational tools to analyze torus bifurcations and breakdowns to demonstrate that they follow up with the existing mathematical theory.  One such reduction tool, especially in the context of neuroscience and neurophysiological models, is to construct reduced Poincar\'e maps from time series. This can be done using the map: $V_n^{\rm min} \to V_{n+1}^{\rm min}$ for minimal or maximal values found in of voltage traces. Using this approach we have shown that the emergence, evolution of invariant circles corresponding to tori in the phase space and slow modulations in voltage traces in our simulation match very well the non-local theory proposed in the classical works by V.S. Afraimovich and L.P. Shilnikov\cite{Annulus1977,some1974,afraimovich1991invariant, afraimovich1974small,AS1983}. Namely, such an invariant circle, born smooth and round, first becomes non-smooth, resonant and next gets distorted by developing homoclinic tangles of unstable sets of resonant saddle periodic orbits on the torus. 
The latter leads to the torus breakdown and to the onset of emergent hyperbolic dynamics.
This prerequisite for deterministic chaos competes with regular dynamics due to the presence of stable periodic orbits overshadowing its host, i.e. the resonant torus. 	Weakly resonant tori exist in narrow ranges in the parameter space of the system. 
The same holds true for ergodic tori as they quickly become non-smooth between resonant zones densely populating the parameter space of such  systems.                         
 
The other option for the visualization  and analysis of tori is to use the return map, $\tau_{n} \to \tau_{n+1}$, for recurrent time intervals between consecutive spikes in the voltage traces. This approach is illustrated in Fig.~\ref{fig:ftTwo}. The map in Fig.~\ref{fig:ftTwo}$\mathbf{A}$ shows the saddle IC (red, $\mathrm{T^{U}}$) enclosing a stable FP (blue, $\mathrm{PO^{S}}$)  corresponding to robust and periodic  tonic spiking  in the Purkinje cell model. 
		The return map in Fig.~\ref{fig:ftTwo}$\mathbf{B}$ depicts the stable IC (blue, $\mathrm{T^{S}}$) encircling an repelling tonic-spiking FP (red, $\mathrm{PO^{U}}$) in the 3F normal-form model~(\ref{eq:NF}). This demonstrates that both amplitude- and  frequency-modulation approaches suite well to study quasi-periodicity using the  observables like recordings of voltage traces.        

 We note that transitions from tonic-spiking to bursting in the given family of slow-fast models can also be underlined by a cascade of period-doubling bifurcations, instead  of the torus bifurcation.
 We proposed and tested the divergence-sign assessment to predict the type of bifurcation. The test works well for 3D model, while its applicability  is problematic for higher dimensional systems, as one has to single out a 3D subspace in restriction to which the analogous divergence should be evaluated.     
 It has become more or less clear that the choice between these two bifurcations is implicitly determined by multiple or just two time-scale reciprocal interactions of dynamic variables of the slow-fast model, as well as by the smoothness of the fold in the averaged system. The smoother such a fold and the closer the system to the singular limit, the more likely it undergoes the torus bifurcation yet unclear weather it will be sub or super-critical.

\section{Acknowledgments} 
This work was partially funded by NSF grant IOS-1455527, RSF grant 14-41-00044 at the Lobachevsky University of Nizhny Novgorod, RFFI grant 11-01-00001, and MESRF project 14.740.11.0919. H.J. and A.S. thank GSU Brain and Behaviors Initiative for the fellowship and pilot grant support. We are grateful to H. Meyer for his patient MATCONT guidance, and to all members of Shilnikov's NeurDS lab for the helpful discussions and suggestions. A.N. acknowledges support from the Quantitative Biology Institute and from Neuroscience Program at Ohio University. 

\bibliography{torus_cite}
\section{Appendix}
The Purkinje cell equations are as follows: 
\begin{widetext}
\begin{align}
\nonumber\mathit{V}' = & -I_{app} -10.0\mathit{n}^{4} \left(\mathit{V}+95.0\right)-152.0\frac{1}{\left(1+e^{-0.1(\mathit{V}+34.5)}\right)^{3}}\mathit{h}(\mathit{V}-50.0)\\
\nonumber&-2.0(\mathit{V}+70.0)-1.0\mathit{c}^{2}(\mathit{V}-125.0)
-0.75\mathit{M}(\mathit{V}+95.0)\\
\nonumber\mathit{h}' = & \frac{1}{\left(1+e^{\frac{\mathit{V}+59.4}{10.7}}\right)\left(0.15+\frac{1.15}{1+e^{\frac{\mathit{V}+33.5}{15.0}}}\right)}(1-\mathit{h})
-\left(1-\frac{1}{1+e^{\frac{\mathit{V}+59.4}{10.7}}}\right)\frac{1}{0.15+\frac{1.15}{1+e^{\frac{\mathit{V}+33.5}{15.0}}}}\mathit{h}\\
\mathit{n}' = & \frac{1}{\left(1+e^{-0.1(\mathit{V}+29.5)}\right)\left(0.25+4.35e^{-0.1\left|\mathit{V}+10.0\right|}\right)}(1-\mathit{n}) \label{eq:purk}
\\
\nonumber&-\left(1-\frac{1}{\left(1+e^{-0.1(\mathit{V}+29.5)}\right)}\right)\frac{1}{0.25+4.35e^{-0.1\left|\mathit{V}+10.0\right|}}\mathit{n}\\
\nonumber\mathit{c}' = & \frac{1.6(1.0-\mathit{c})}{1+e^{-0.072(\mathit{V}-5.0)}}-\frac{0.02\mathit{c}(\mathit{V}+8.9)}{-1+e^{0.2(\mathit{V}+8.9)}}\\
\nonumber\mathit{M}'= & \frac{0.02}{1+e^{-0.2(V+20.0+\Delta)}}(1-\mathit{M})-0.01e^{\frac{-(\mathit{V}+43.0)}{18.0}}\mathit{M}.
\end{align}
\end{widetext}

\subsection*{Matlab codes}
1. 12D hair cell model is given by 
\begin{lstlisting}
gL=0.175;
gk1=29.5;
gh=2.2;
gCa=1.2;
BK=0.01;
Ek=-95.e-3;
Eh=-45.e-3;
farad=96485.0;
const1=3793550.768;
const2=39.317;
Ki=0.112;
Ko=0.002;
Eca=42.e-3;
km1=300.0;
km2=5000.0;
km3=1500.0;
k01=6.e-6;
k02=45.e-6;
k03=20.e-6; 
ddel1=0.2;
alphc0=450.0; 
VA=33.e-3;
betc=2500.0;
Ks=2800.0; 
Cvol=1.25e-12;
eeps=3.4e-5;
z=2;
U=0.005;
Cm=10.e-12;
PBKS=2.e-13;
PBKT=14.e-13;
Pdrk=2.4e-14;

dy = zeros(12,1);

mk1sinf=1.0/(1.0+exp((y(1)+0.110)/0.011));
tk1f=0.7e-3*exp((y(1)+0.120)/(-0.0438))+0.04e-3;
tk1s=14.1e-3*exp((y(1)+0.120)/(-0.028))+0.04e-3;
Ik1=gk1*1e-9*(y(1)-Ek)*(0.7*y(2)+0.3*y(3));
Ih=gh*1e-9*(y(1)-Eh)*(3*y(4)*y(4)*(1.0-y(4))+y(4)*y(4)*y(4));
mhinf=1.0/(1+exp((y(1)+0.087+Iscan)/0.0167));
th=0.0637+0.1357*exp(-((y(1)+0.0914)/0.0212)*((y(1)+0.0914)/0.0212));
ex=exp(-const2*y(1));
permiab=const1*y(1)*(Ki-Ko*ex)/(1.0-ex);
mdrkinf=1.0/sqrt(1.0+exp(-(y(1)+0.0483)/0.00419));
alfdrk=1.0/(0.0032*exp(-y(1)/0.0209)+0.003);
betdrk=1.0/(1.467*exp(y(1)/0.00596)+0.009);
taudrk=1.0/(alfdrk+betdrk);
Idrk=y(5)*y(5)*Pdrk*permiab;
mCainf=1.0/(1.0+exp(-(y(1)+0.055)/0.0122));
tauCa=0.046e-3+0.325e-3*exp(-((y(1)+0.077)/0.05167)*((y(1)+0.077)/0.05167));
Ica=gCa*1e-9*(y(1)-Eca)*y(6)*y(6)*y(6);
ex1=exp(ddel1*z*const2*y(1));
k1=km1/k01*ex1;
k2=km2/k02;
k3=km3/k03*ex1;
alphc=alphc0*exp(y(1)/VA);
yy7=1.0-(y(7)+y(8)+y(9)+y(10));
hbktinf=1.0/(1.0+exp((y(1)+61.6e-3)/3.65e-3));
taubkt=2.1e-3+9.4e-3*exp(-((y(1)+66.9e-3)/17.7e-3)*(y(1)+66.9e-3)/17.7e-3);
IBKS=PBKS*BK*permiab*(y(9)+y(10));
IBKT=PBKT*BK*permiab*(y(9)+y(10))*y(12);
IL=gL*1e-9*y(1);
dy(1)=-(Ik1+Ih+Idrk+Ica+IBKS+IBKT+IL)/Cm;
dy(2)=(mk1sinf-y(2))/tk1f;
dy(3)=(mk1sinf-y(3))/tk1s;
dy(4)=(mhinf-y(4))/th;
dy(5)=(mdrkinf-y(5))/taudrk;
dy(6)=(mCainf-y(6))/tauCa;
dy(7)=k1*y(11)*yy7+km2*y(8)-(km1+k2*y(11))*y(7);
dy(8)=k2*y(11)*y(7)+alphc*y(9)-(km2+betc)*y(8);
dy(9)=betc*y(8)+km3*y(10)-(alphc+k3*y(11))*y(9);
dy(10)=k3*y(11)*y(9)-km3*y(10);
dy(11)=-U*Ica/(z*farad*Cvol*eeps)-Ks*y(11);
dy(12)=(hbktinf-y(12))/taubkt;
\end{lstlisting}

2. 5D Purkinje Cell model is given by 
\begin{lstlisting}
V'=-Iapp-(95.0+V)*(0.75*M)-1.0*c^2*(-125.0+V)-2.0*(70.0+V)-152*(1.0/(1.0+exp((-V-34.5)/10.0)))^3.0*h*(-50.0+V)-10.0*n^4.0*(95.0+V);
h'=1.0/(1.0+exp((V+59.4)/10.7))/(0.15+1.15/(1.0+exp((V+33.5)/15.0)))*(1.0-h)-(1.0-1.0/(1.0+exp((V+59.4)/10.7)))/(0.15+1.15/(1.0+exp((V+33.5)/15.0)))*h;
n'=1.0/(1.0+exp((-V-29.5)/10.0))/(0.25+4.35*exp(-abs(V+10.0)/10.0))*(1.0-n)-(1.0-1.0/(1.0+exp((-V-29.5)/10.0)))/(0.25+4.35*exp(-abs(V+10.0)/10.0))*n;
c'=(1.6*(1.0-c))/(1.0+exp(-0.072*(-5.0+V)))-(0.02*c*(8.9+V))/(-1.0+exp((8.9+V)/5.0));
M'=0.02/(1.0+exp((-20.0-shift-V)/5.0))*(1.0-M)-0.01*exp((-43.0-V)/18.0)*M;
\end{lstlisting}

3. 4D parabolic burster model is given by 
\begin{lstlisting}
V1'=(-gL*(V1-EL)-(gK*n1+gKCa*(Ca1/(1+Ca1)))*(V1-EK)-(gCa*(0.5*(1+tanh((V1+1.2)/18)))+gCaS*x1)*(V1-ECa)+Iapp)/Cm;
n1'=phi*cosh((V1-12)/34.8)*((0.5*(1+tanh((V1-12)/17.4)))-n1);
x1'=eps*((0.5*(1+tanh((V1-12)/24)))-x1)/0.05;
Ca1'=eps*(-mu*((gCa*(0.5*(1+tanh((V1+shift+1.2)/18)))+gCaS*x1)*(V1+Cashift-ECa))-Ca1);
parameters: gL=2, gK=8, gKCa=1; gCa=4, gCaS=1, EL=-60, EK=-84, ECa=120, phi=0.06666, eps=0.0005, mu=0.025, Iapp=68.26, Cm=20.
\end{lstlisting}

\end{document}